\documentclass[prd, 11pt,nofootinbib, superscriptaddress, preprintnumbers,floatfix]{revtex4}
\usepackage[utf8]{inputenc}
\usepackage{amsmath,amssymb}
\usepackage{graphicx}
\usepackage{graphics}
\usepackage{subfigure}
\usepackage{pdfpages}
\usepackage{pdflscape}
\usepackage{rotating}
\usepackage{hyperref}
\usepackage{nicefrac}
\usepackage{xr}
\usepackage{booktabs}
\usepackage{import}
\usepackage{braket}
\usepackage{multirow}
\usepackage{here}
\usepackage{color}
\usepackage{dsfont}
\usepackage{placeins}

\def\ProvideLabels#1{}

\graphicspath{{plots/}}

\begin{document}

\title{Vector particle scattering on the lattice}
%\preprint{arXiv:XXXX.YYYYY}

\newcommand\bn{HISKP and BCTP, Rheinische Friedrich-Wilhelms Universit\"at Bonn, 53115 Bonn, Germany}

\author{F.~Romero-López}\affiliation{\bn}
\author{A.~Rusetsky}\affiliation{\bn}
\author{C.~Urbach}\affiliation{\bn}

\begin{abstract}
 %In this work we present an extension of the Lüscher formalism to include the interaction of particles with spin, focusing on the scattering of two vector particles. The derived formalism will be applied to scalar QED in the Higgs Phase, where the $U(1)$ gauge boson acquires mass.
In this work, we present an explicit form of the L\"uscher equation and consider the 
construction of the operators in different irreducible representations
for the case of scattering of two vector particles. The formalism
is applied to scalar QED in the Higgs Phase, where the $U(1)$ gauge boson acquires
mass.
\end{abstract}

\maketitle

%\clearpage

%\input{intro1}

\section{Introduction}

The study of scattering in Lattice Field Theory (LFT) starts with the original work
of L\"uscher~\cite{Luscher}. In this work, an equation that relates the
scattering phase shift of a spinless particle to the finite-volume spectrum in the
rest frame was derived. The formalism has been extended to moving frames~\cite{Gottt},
$\pi-N$ scattering ~\cite{Bernard:2008ax}, $N-N$ scattering~\cite{Briceno:2013lba},
different masses~\cite{Bernard:2012bi,ZiwenFu}, moving frames with different
masses~\cite{Leskovec:2012gb,Gockeler:2012yj} and any multichannel system with
arbitrary spin, momentum and masses~\cite{Briceno:2014oea}. In practice, the
extraction of the phase shifts from data in case of coupled channels is most
conveniently done by the use of the $K$-matrix approach. This method was first proposed
in Ref.~\cite{Lage:2009zv}, and a complete description can be found, e.g., in the recent
work~\cite{MORNINGSTAR2017477}.

For the case of scattering of vector particles, there may be interesting issues that can be addressed through LFT, such as the possibility of the Higgs boson to be a bound state of two $W$ bosons. This is the case for a model proposed in Refs.~\cite{Frezzotti:2014wja, Frezzotti:2016bes}, where a ``superstrong interaction'' together with superstrongly interacting particles are present.
Instead of the Higgs mechanism, a nonperturbative mass generation mechanism is suggested.
The model is strongly coupled at the relevant scale, and, therefore, LFT is the approach to test whether this mechanism exists or not
(see Refs.~\cite{Capitani:2017trq,Capitani:2017ucw} for a first numerical investigation of this model).
One possible consequence of this model could be that the Higgs represents a bound state in the $WW$ channel. This justifies a thorough study of the $WW$ interactions (including both the bound spectrum and scattering) within LFT, which is possible by using L\"uscher's approach.

The aim of the present work is to study the vector-vector scattering process in a toy model,
which is the first step towards applying the same method in physically more interesting
cases. To this end, we rederive the L\"uscher equation for scattering
of particles with arbitrary spin by using nonrelativistic effective theory and check that
the results obtained are in agreement with Ref.~\cite{Briceno:2014oea}.
We will further focus on the case of two identical vector particles and we make use
of the spatial symmetries of the lattice to factorize the L\"uscher equation.
We explicitly construct the operators that transform under a certain irreducible
representation of the
spatial symmetry group, and, using these, we gain access to the different phase
shifts of the theory.
The approach will be tested in scalar QED, for which numerical results will be shown.
For a first account of this work we refer to Ref.~\cite{Romero-Lopez:2017gag}.

\section{Scattering of two vector particles}\label{sec-1}
\subsection{Derivation of L\"uscher equation for arbitrary spin}

Let us consider a system of two particles with masses $m_i$, $i=1,2$ in $d=3$
dimensions.  The system is described by the effective nonrelativistic Lagrangian
\begin{equation}
 \mathcal{L} = \phi_1^\dagger 2 W_1(i\partial_t - W_1)\phi_1 + \phi_2^\dagger 2 W_2(i\partial_t - W_2)\phi_2 + \mathcal{L}_I.
\end{equation}
Here, $\phi_i$ are the nonrelativistic fields with spin $s_i$, $W_i = (m_i^2 - \nabla^2)^{1/2}$ and the interactions are contained in ${\mathcal{L}}_I$. The corresponding nonrelativistic propagators, with $\omega_i({\mathbf p}) = (m_i^2 + \mathbf{p}^2)^{1/2}$, are diagonal in the spin indices $\nu,~\nu'$:
\begin{equation}
\bigl( S_i(p)\bigr)_{\nu'\nu} = \frac{1}{2\omega_i(\mathbf{p})} \frac{1}{\omega_i(\mathbf{p})-p^0-i \epsilon} \delta_{\nu'\nu} ,\quad\quad
\nu',\nu=1,\cdots,2s_i+1\, .
\end{equation}
The scattering T-matrix is defined through the Lippman-Schwinger (LS) equation
\begin{equation}
 T(z) = (-H_I) + (-H_I)(-G_0(z)) T(z),
 \label{eq:LS}
\end{equation}
where $H_0$ and $H_I$ are obtained from the Lagrangian in the usual way and
$G_0(z) = (z-H_0)^{-1}$ is the free resolvent. The two-particle states 
with a total spin $S$ are given by
\begin{equation}
  \ket{\mathbf{k_1},\mathbf{k_2},S,\nu} \equiv \ket{\mathbf P,\mathbf k,S,\nu}\,,
  \label{eq:statesLab}
\end{equation}
with normalization
\begin{equation}
  \braket{\mathbf{P'},\mathbf{k'},S',\nu'|\mathbf P,\mathbf k,S,\nu  } =
  4 \omega_1({\mathbf{k_1}}) \omega_2({\mathbf{k_2}}) (2\pi)^d \delta^d(\mathbf{P'}-\mathbf P) (2\pi)^d\delta^d(\mathbf{k'}-\mathbf k) \delta_{S'S} \delta_{\nu'\nu}\,, 
\end{equation}
where ${\mathbf{k_1}}, {\mathbf{k_2}}$ are the momenta of the particles,
$S$ and $\nu$ denote the total spin and its projection for the two-particle system, respectively, and
$\mathbf P$, $\mathbf k$ are the total and relative momenta in the laboratory frame:
\begin{eqnarray}
  \mathbf P = \mathbf {k_1}+\mathbf{k_2},&&\quad 
  \mathbf k = \mu_2 \mathbf{k_1} - \mu_1 \mathbf{k_2}, \quad\quad
  \mu_{1,2} = \frac{1}{2}\left(1\pm \frac{m_1^2- m_2^2}{ P^2}\right), 
  \nonumber\\[2mm] 
  P_0=\omega_1({\mathbf{k_1}}) +\omega_2({\mathbf{k_2}})\, ,&&\quad 
  P^2=P_0^2- \mathbf{P^2}\,. 
\end{eqnarray}
Now define the matrix elements:
\begin{eqnarray}
 t^{S'S}_{\nu'\nu}(\mathbf{k'},\mathbf k,\mathbf P,z) &=&\int \frac{d^d\mathbf {P'}}{(2\pi)^d} \braket{\mathbf{P'},\mathbf{k'},S',\nu'|T(z)|\mathbf P,\mathbf k,S,\nu}, \label{eq:tnunu}
\\[2mm]
 h^{S'S}_{\nu'\nu}(\mathbf{k'},\mathbf k,\mathbf P) &=& \int \frac{d^d\mathbf{P'}}{(2\pi)^d} \braket{\mathbf{P'},\mathbf{k'},S',\nu'|(-H_I)|\mathbf P,\mathbf k,S,\nu}. \label{eq:hnunu}
\end{eqnarray}
One may rewrite the LS equation in terms of matrix elements, using Eqs.~(\ref{eq:tnunu}) and (\ref{eq:hnunu}):
\begin{eqnarray}
 t^{S'S}_{\nu'\nu}(\mathbf{k'},\mathbf k,\mathbf P,z) 
&=& h^{S'S}_{\nu'\nu}(\mathbf {k'},\mathbf k,\mathbf P) 
\nonumber\\[2mm] 
&+& \int \frac{d^d\mathbf{q}}{(2\pi)^d}\sum_{S''\nu''}  \frac{  h^{S'S''}_{\nu'\nu''}(\mathbf{k'},\mathbf{q},\mathbf P) t^{S''S}_{\nu''\nu}(\mathbf{q},\mathbf k,\mathbf P,z)}{4\omega_1(\mathbf{q_1})\omega_2(\mathbf P-\mathbf{q_1})(\omega_1(\mathbf{q_1})+\omega_2(\mathbf{P}-\mathbf{q_1})-z)}, \label{eq:LS2}
\end{eqnarray}
where we define $\mathbf q = \mu_2 \mathbf{q_1} - \mu_1 \mathbf{q_2} $, as in 
Eq.~(\ref{eq:statesLab}). A key point here is that the elementary bubble 
(the free two-particle propagator, integrated over the relative momentum)
is diagonal in spin, because also the single particle propagators are. However, the scattering amplitude need not be diagonal.

Now define the projectors to the partial waves in the CM frame, whose momenta are $\mathbf{k^*}$:
\begin{eqnarray}
 \Pi_{\nu'\nu}^{A'A} ( \mathbf {k'^*}, \mathbf{ k^*}) =\sum_{\rho, \rho'} U^{(S')}_{\nu'\rho'}(\mathbf{k'^*})^* U^{{(S)}}_{\nu \rho}(\mathbf{k^*}) (\mathcal{Y}_{J'l'S'\mu'}(\mathbf{k'^*},\rho'))^* \mathcal{Y}_{JlS\mu}(\mathbf{ k^*},\rho)\, , 
 \end{eqnarray}
where $A=(J,l,S,\mu)$,  $A'=(J',l',S',\mu')$ represent multi-indices and  $U^{(S)}_{\nu \rho}(\mathbf{k^*})$ is the unitary transformation of the spin indices under a boost. The spherical harmonics with spin 
are defined as
\begin{equation}
 \mathcal{Y}_{JlS\mu}(\mathbf k,\nu) = \sum_{m,\sigma} \braket{l S m  \sigma | J\mu} |\mathbf k|^l Y_{lm}(\mathbf{\hat k}) \chi^S_{\sigma}(\nu) \equiv |\mathbf k|^l Y_{JlS\mu}(\mathbf{\hat k},\nu)\, ,\quad\quad 
\mathbf{\hat k}= \mathbf k / |\mathbf{k}|\, ,
\label{eq:sphar}
\end{equation}
where $Y_{lm}$ denote usual spherical harmonics.

Using the projectors, the quantities in Eqs.~(\ref{eq:tnunu}) and (\ref{eq:hnunu}) can be expanded as
\begin{eqnarray}
t^{S'S}_{\nu'\nu}(\mathbf{k'},\mathbf k,\mathbf P,z) &=& 4\pi \sum_{\substack{J'l'\mu', Jl\mu}}
 \Pi_{\nu'\nu}^{A'A}(\mathbf{k'^*},\mathbf{k^*}) t_{A'A}(|\mathbf{k'^*}|,|\mathbf{k^*}|,\mathbf P,z),\label{eq:texpansion} \\[2mm]
h^{S'S}_{\nu'\nu}(\mathbf{k'},\mathbf{k},\mathbf P) &=& 4\pi \sum_{J'l'\mu',Jl\mu} \Pi_{\nu'\nu}^{A'A}(\mathbf{k'^*},\mathbf{k^*}) h_{A'A}(|\mathbf{k'^*}|,|\mathbf{k^*}|, \mathbf P).\label{eq:hexpansion}
\end{eqnarray}
If the system is placed in a box of a size $L$, the momenta are quantized.
The integral in the LS equation should be replaced by a sum:
\begin{equation}
 \int \frac{d^d\mathbf {q}}{(2\pi)^d}  \rightarrow \frac{1}{L^3}\sum_{\mathbf{q}}\,,
\quad\quad \mathbf{q}=\frac{2\pi}{L}(\mathbf{n}-\mu_1\mathbf{d})\, ,\quad
\mathbf{n}\in\mathbb{Z}^3\, ,
 \label{eq:sumint}
\end{equation}
where $2\pi\,\mathbf{d}/L=\mathbf{P}$.

By plugging the Eqs.~(\ref{eq:texpansion}) and (\ref{eq:hexpansion}) into the 
finite volume equivalent of Eq.~(\ref{eq:LS2}), one gets:
\begin{eqnarray}
 t_{A'A}(s,\mathbf P) - h_{A'A}(s,\mathbf P) = \frac{k^*}{8\pi \sqrt{s}}\sum_{B',B} h_{A'B'}(s,\mathbf P)( (\mathbf{k^*})^{l+l'} i^{l-l'} \delta_{S_{B'} S_B} \mathcal{M}_{B'B}(s,\mathbf P)) t_{BA}(s,\mathbf P), 
 \label{eq:LS3}
\end{eqnarray}
with $s=P^2$ and $S_B$ being the spin of the multi-index $B$. Note that using dimensional regularization, one is able to rewrite
the LS equation as an algebraic equation, involving only the on-shell quantities. Hence, the quantities $t_{A'A}$ and $h_{A'A}$ in Eq.~(\ref{eq:LS3})
coincide with their counterparts from Eqs.~(\ref{eq:texpansion}) and (\ref{eq:hexpansion}) on shell, i.e.,
\begin{equation}
 |\mathbf{k'^*}|=|\mathbf{k^*}|=\frac{\lambda^{1/2}(s,m_1^2,m_2^2)}{2\sqrt{s}}\, ,\quad\quad z=P_0\, 
\end{equation}
where $\lambda$ denotes the triangle function.

Now, using unitarity of the transformation of the spin indices, one arrives at
\begin{eqnarray}
 \mathcal{M}_{J'l'S'\mu',JlS\mu}(s,\mathbf P) 
= \frac{32\pi^2}{|\mathbf{k^*}|} \frac{\sqrt{s}}{L^3} 
i^{l-l'} \delta_{S'S}\sum_\nu \sum_{\bf q} 
\frac{({Y}_{J'l'S\mu'}(\mathbf{\hat q^*},\nu))^*{Y}_{JlS\mu}(\mathbf{\hat q^*},\nu)
}{4\omega_1(\mathbf{q_1})\omega_2(\mathbf P-\mathbf{q_1}) 
(\omega_1(\mathbf{q_1})+\omega_2(\mathbf P-\mathbf{q_1})-P_0)}\, .
\nonumber\\
\end{eqnarray}
This matrix can be related to its equivalent for scalar particles by using Eq.~(\ref{eq:sphar}):
\begin{equation}
 \mathcal{M}_{J'l'S'\mu',JlS\mu}= \delta_{S'S} \sum_{m',m,\sigma} 
\braket{ l'Sm'\sigma | J' \mu'} \braket{ lSm\sigma | J \mu} 
\mathcal{M}_{l'm',lm},
\end{equation}
where we used the identity \cite{Gasser:2011ju,Bernard:2012bi} (with $\mathbf q = \mathbf{q_1} - \mu_1 \mathbf P$)
\begin{eqnarray}
 \frac{1}{4\omega_1\omega_2 (\omega_1+\omega_2-P_0)} &=& \frac{1}{2P_0} \frac{1}{\mathbf q^2-\frac{(\mathbf q \mathbf P)^2}{P_0^2}-(\mathbf{k^*})^2}  
\nonumber\\[2mm]
&+& \frac{1}{4\omega_1\omega_2}\left( \frac{1}{\omega_1+\omega_2+P_0} - \frac{1}{\omega_1-\omega_2+P_0} - \frac{1}{\omega_2-\omega_1+P_0}   \right),
\label{eq:iden}
\end{eqnarray}
kept only the singular part [first term in Eq.~(\ref{eq:iden})] and used $(\mathbf{q^*})^2= \mathbf{q}^2 - \frac{(\mathbf{qP})^2}{P^2_0}$.
This way, and up to exponentially suppressed terms,
% ($|\mathbf{q^*}| \rightarrow |\mathbf{k^*}|$), 
$\mathcal{M}_{l'm',lm}$ is given by (see Ref.~\cite{Bernard:2012bi})
\begin{equation}
 \mathcal{M}_{l'm',lm}(\mathbf{k^*},s) = \frac{(-1)^{l'}}{\pi^{3/2}\gamma} 
\sum_{j=|l-l'|}^{l+l'}\sum_{s=-j}^j \frac{i^j}{\eta^{j+1}} Z^d_{js}(1,s)^* C_{l'm',js,lm},  \ \ \ \ \ \eta= \frac{|\mathbf{k^*}|L}{2\pi}, 
\end{equation}
where
\begin{equation}
 C_{l'm',js,lm} = (-1)^{m} i^{l'-j+l} \sqrt{(2l+1)(2l'+1)(2j+1)}
 \begin{pmatrix}
 l' & j & l \\ m' & s & -m
 \end{pmatrix}
 \begin{pmatrix}
 l' & j & l \\ 0 & 0 & 0
 \end{pmatrix},
\end{equation}
\begin{equation}
 Z^d_{lm}(1,s) = \sum_{\mathbf r\in P_d} \frac{|\mathbf r|^l Y_{lm}(r)}{\mathbf r^2-\eta^2}, \ \ \ P_d=\{\mathbf{r_{||}}= \gamma^{-1}(\mathbf{n_{||}}-\mu_1 \mathbf d), \mathbf{r_{\perp}}= \mathbf{n_{\perp}}   \}\,,
 \label{eq:defZ}
\end{equation}
$\gamma=(1-\mathbf{P}^2/P_0^2)^{-1/2} $ and $\mathbf{n} \in \mathbb{Z}^3$. One can see that Eq.~(\ref{eq:LS3}) is a matrix equation, and the poles in $t_{A'A}$ arise when
\textcolor{black}{
\begin{equation}
  \det\,{\cal A}=0\, ,
  \end{equation}
  where ${\cal A}$ is a matrix
\textcolor{black}{
  \begin{equation}
 {\cal A}_{J'l'S'\mu',JlS\mu}=   \frac{8\pi \sqrt{s}}{|\mathbf{k^*}|^{l+l'+1}} (h^J_{l'S',lS})^{-1} \delta_{J'J} \delta_{\mu'\mu}-  \delta_{S'S} \mathcal{M}_{J'l'S\mu',JlS\mu}\, .
 \label{eq:luscher1}
\end{equation}
}
Here it is already implied that $J$ and $\mu$ are conserved in scattering processes in the infinite volume, i.e. $h_{J'l'S'\mu',JlS\mu} = h^J_{l'S',lS}\delta_{J'J}\delta_{\mu'\mu}$, and the factor $i^{l-l'}$ can be dropped in the determinant.}

Now, in order to express this equation in a more compact way, one uses the standard definition of the $S$ matrix (see \cite{Briceno:2013bda}, for nucleon-nucleon scattering), $S = e^{2i\delta(s)}$, in terms of the phase shift $\delta(s)$. This way, one can write down $h^J_{l'S',lS}$ in terms of $\delta$:
\begin{equation}
 h^J_{l'S',lS} = \frac{8\pi\sqrt{s}}{|\mathbf{k^*}|^{l+l'+1}} (\tan \delta)^J_{l'S',lS}
\, . \label{eq:hAA}
\end{equation}
Plugging it in Eq.~(\ref{eq:luscher1}), we arrive at
\begin{equation}
{\cal A}_{J'l'S'\mu',JlS\mu}=  (\cot \delta)^J_{l'S',lS} \delta_{J'J}\delta_{\mu'\mu}-\delta_{S'S} \mathcal{M}_{J'l'S\mu',JlS\mu}\, .
 \label{eq:luscher2}
\end{equation}

\subsection{Two vector Particles \label{sec:twovector}}

A system of two identical vector particles can couple to total spin $S=0,1,2$. Even spin combinations are symmetric under the exchange of two particles, whereas odd combinations are antisymmetric. The same holds for the angular momentum $L$. The possible combinations of $S$ and $L$ to $J^P$, respecting Bose statistics (totally symmetric state), are listed in the Table \ref{tab:Jp}. The combinations that have mixing are in the same column in the table and correspond to same $J^P$ but different $L$, $S$.
%\begin{center}
\begin{table}[H]
\centering
\begin{tabular}{| c | c | c | c | c | c | c | }
\hline
 $J^P$                     &$0^+$    & $0^-$   & $1^+$   & $1^-$   & $2^+$   & $2^-$\\ \hline
 \multirow{3}{*}{$\{S,L\}$}&$\{0,0\}$&         &         &         &$\{0,2\}$& \\ \cline{2-7}
                           &         &$\{1,1\}$&         &$\{1,1\}$&         &$\{1,1\}$,$\{1,3\}$ \\ \cline{2-7}
                           &$\{2,2\}$&         &$\{2,2\}$&         &$\{2,0\}$,$\{2,2\}$,$\{2,4\}$&\\ \cline{1-7}
\end{tabular}  
\caption{Possible values of $J^P$ with $J<3$. \label{tab:Jp}}
\end{table}
%\end{center}
The possible mixings can be parametrized by a mixing angle and two eigenvalues. This would be analogous to the parametrization of the mixings for two nucleons in Ref. \cite{Briceno:2013bda}; for example:
\begin{equation}
 \cot \delta ^{0^+} = \begin{pmatrix} \cos \epsilon_0&  -\sin \epsilon_0 \\ \sin \epsilon_0& \cos \epsilon_0  \end{pmatrix}
 \begin{pmatrix} \cot \delta_{1}^{0^+}& 0 \\ 0&  \cot \delta_{2}^{0^+} \end{pmatrix}
 \begin{pmatrix} \cos \epsilon_0& \sin \epsilon_0 \\ -\sin \epsilon_0& \cos \epsilon_0  \end{pmatrix}, \label{eq:paramcot}
\end{equation}
Since no mixing occurs between even and odd spins, neither in the $\mathcal{M}$ matrix, nor in the phase shifts, Eq.~(\ref{eq:luscher2}) factorizes for even and odd spin.

%(\theta_{02},\theta_{04},\theta_{24})

%\clearpage

\subsection{Effective range expansion in case of multiple channels \label{sec:effrange}}

For the scattering of two spinless particles, it is well known (see Ref. \cite{Bethe}) that the phase shift can be parametrized as a polynomial of $\mathbf k^2$:
\begin{equation}
 \mathbf k^{2l+1} \cot \delta_l = \sum_{n=0} a_{nl} \mathbf k^{2n}.
\end{equation}
One obviously needs an analog of this parametrization in the multichannel case as well \footnote{ An equivalent derivation can be found in Refs. \cite{ROSS1961147} and \cite{MORNINGSTAR2017477}.}.
In order to derive such a parametrization, we note that, within the effective field theory, the left-hand side of Eq.~(\ref{eq:hexpansion}) has a Taylor expansion in momenta. Taking now into account the fact that the projector on the right-hand side of the same equation contains the factor $|\mathbf{k}^*|^{l+l'}$, from Eq.~(\ref{eq:hAA}) one may finally conclude that, on the mass shell,
\begin{equation}
 \mathbf k^{l+l'+1} \cot \delta^J_{l'S',lS} = \sum_{n=0} (a_n)_{l'S',lS} \mathbf k^{2n}. \label{eq:EREMC}
\end{equation}

\subsection{Reduction of the L\"uscher equation}

Our aim here is to construct the basis
vectors of all irreducible representations (irreps) from the basis vectors of the irreps
of the rotation group, corresponding to the symmetry in the infinite
volume, and to (partially) diagonalize the L\"uscher equation in this
new basis. The general procedure is well known in the literature, so
we shall skip many details\footnote{See, for example, Ref.~\cite{Thomas:2011rh},
  where the same problem has been considered by using the helicity formalism.}.
Let $\mathcal{G}$ be a full octahedral group including inversions, or  a subgroup thereof (little group), 
which is the symmetry group in the moving frames. 
Let $\Gamma$ be a certain
irrep of $\mathcal{G}$, and let $\alpha=1,\ldots\mbox{dim}\,\Gamma$ be an index labeling basis vector in this representation. One can construct these basis vectors by applying
certain projection operators to the basis vectors of the irreps of the rotation group. 
These (unnormalized) projectors are given by
\begin{equation}
( P^{\Gamma,J,l}_{\alpha\beta})_{\mu\mu'} = \sum_{\mathcal{S} \in G} (R_{\alpha\beta}^\Gamma(\mathcal{S}))^* { D}^J_{\mu\mu'}(\mathcal{S})\, .
\end{equation}
Here, $D^J_{\mu\mu'}(\mathcal{S})$ denotes the usual Wigner matrix, if $\mathcal{S}$ corresponds to a pure rotation. Otherwise, the group elements can be represented as
$\mathcal{S}=I\bar{\mathcal{S}}$, where $I$ is an inversion and $\bar{\mathcal{S}}$
is a pure rotation. In this case, we define $D^J_{\mu\mu'}(\mathcal{S})=(-1)^lD^J_{\mu\mu'}(\bar{\mathcal{S}})$. 
 Furthermore, $R_{\alpha\beta}^\Gamma(\mathcal{S})$ denotes a matrix representation of
$\mathcal{G}$ in the irrep $\Gamma$.

These projectors must be applied to the basis vectors of the irreps of the rotation
group $\ket{J,S,l,\mu}$ with indices $\beta$ and $\mu$ fixed
\begin{equation}
 \ket{\Gamma, \alpha, J, S, l, n} \propto \sum_{\mu'} ( P^{\Gamma,J,l}_{\alpha\beta})_{\mu\mu'} \ket{J,S,l,\mu'},
\label{eq:proj}
\end{equation}
%\begin{equation}
% (e^{\Gamma J S l \beta}_{\alpha})_\mu \propto \sum_{\mu'} ( P^{\Gamma,J}_{\alpha\beta})_{\mu\mu'} \phi_{\mu'}.
%\end{equation}
where $n$ labels the number of multiple occurrences of $\Gamma$.
The different spatial symmetry groups, with their irreducible representations and the corresponding elements are listed in Appendices \ref{app:convention} and \ref{app:grouptables}.

As seen from Eq.~(\ref{eq:proj}),
the basis vectors of the irreducible representations of the symmetry group of the lattice can be expressed in terms of the one of the continuum:
\begin{equation}
 \ket{\Gamma,\alpha,J,l,S,n} = \sum_{\mu} c^{\Gamma n \alpha}_{Jl\mu} \ket{JlS\mu},
\end{equation}
where the Clebsch-Gordan coefficients $c^{\Gamma n \alpha}_{Jl\mu}$ can be 
read from Tables \ref{tab:BV1d} to \ref{tab:BV2c} in Appendix \ref{app:BV}. They are in agreement with those of Refs. \cite{Gockeler:2012yj,Bernard:2008ax}
and obey the usual orthogonality conditions
\begin{equation}
 \sum_{\mu} (c^{\Gamma' n' \alpha'}_{Jl\mu})^*  c^{\Gamma n \alpha}_{Jl\mu} = \delta_{\Gamma' \Gamma} \delta_{\alpha' \alpha} \delta_{n' n}\, .
\label{eq:ortho}
\end{equation}
The matrix $\mathcal{M}$ can be partially diagonalized in the new basis:
\begin{eqnarray}
\braket{\Gamma',\alpha',J',l',S,n'| \mathcal{M}  |\Gamma,\alpha,J,l,S,n} = \mathcal{M}^{\Gamma}_{J'l'Sn',JlSn} \delta_{\Gamma'\Gamma} \delta_{\alpha' \alpha}\, ,
\end{eqnarray}
where
\begin{eqnarray}
\mathcal{M}^{\Gamma}_{J'l'Sn',JlSn} = \sum_{\mu\mu'} ( c^{\Gamma n' \alpha}_{J'l'\mu'})^*  c^{\Gamma n \alpha}_{Jl\mu} 
\mathcal{M}_{J'l'S\mu',JLS\mu}\, ,
\end{eqnarray}
(for a given $\Gamma$ and $\alpha$).
Moreover, the matrix $\cot \delta$ should be written down in
 the same basis as $\mathcal{M}$:
\begin{equation}
(\cot \delta)^{\Gamma}_{J'l'S'n',JlSn}=\sum_{\mu'\mu} ( c^{\Gamma n' \alpha}_{J'l'\mu'})^*c^{\Gamma n \alpha}_{Jl\mu} (\cot \delta)^J_{l'S',lS}\delta_{\mu\mu'}\delta_{JJ'}  =
\delta_{JJ'}\delta_{nn'}(\cot \delta)^J_{l'S',lS}.
\end{equation}
Here, we have used Eq.~(\ref{eq:ortho}) and the fact that only states with 
the same parity can mix. Now one sees that 
the determinant factorizes:

\begin{equation}
  \prod_{S= \substack{even\\odd}}  \ \  \prod_{\Gamma} \ \det {\cal A}^\Gamma  =0\, ,
\end{equation}
where
\begin{equation}
  {\cal A}^\Gamma_{J'l'S'n',JLSn}=
  (\cot \delta)^J_{l'S',lS}\delta_{JJ'}\delta_{nn'}-\delta_{SS'}\mathcal{M}^{\Gamma}_{J'l'Sn',JlSn}
  \, .
\end{equation}
For simplicity, $\mathcal{M}^\Gamma$ will be expressed in terms of functions $\omega_{js}$
\begin{equation}
  \omega_{js} = \frac{\gamma^{-1}\eta^{-j-1}}{\pi^{3/2}\sqrt{2j+1}} Z^d_{js}(1,s).
\end{equation}
Understanding and using some symmetry properties of $\omega_{lm}$ simplifies the expressions of $\mathcal{M}^\Gamma$. A frame-independent property is
\begin{equation}
 Z^d_{lm} = (-1)^m Z_{l-m}^*,
\end{equation}
and it is a direct consequence of the properties of the spherical harmonics.
Moreover, if one changes the order of the particles ($m_1 \leftrightarrow m_2$):
\begin{equation}
 Z^{d,(m_1,m_2)}_{js} = (-1)^j Z^{d,(m_2,m_1)}_{js},
\end{equation}
which for the case of equal masses implies vanishing $Z^d_{js}$ for odd j. Additional symmetry properties of $\omega_{js}$ in the rest frame are listed in Table \ref{tab:symmetryw1} for the rest frame (See also \cite{LUSCHER1991531}) and in Ref. \cite{Gockeler:2012yj} for moving frames. In Appendix \ref{app:M} we give some examples of $\mathcal{M}^\Gamma$ in the rest frame.
\begin{table}[H]
  \centering
  \begin{tabular*}{.8\textwidth}{@{\extracolsep{\fill}}cc}
    \hline\hline
    $l$& $\mathbf{\omega}_l$                                                                                             \\ \hline
    0& $(\omega_{00})$                                                                                                  \\%\hline
    1& $(0,0,0)$                                                                                                        \\%\hline
    2& $(0,0,0,0,0)$                                                                                                    \\%\hline
    3& $(0,0,0,0,0,0,0)$                                                                                               \\%\hline
    4& $(\frac{5}{\sqrt{70}}\omega_{40},0,0,0,\omega_{40},0,0,0,\frac{5}{\sqrt{70}} \omega_{40})$ \\ 
    5& $(0,0,0,0,0,0,0,0,0,0,0)$                                                                                               \\%\hline
    6& $(0,0,-\sqrt{\frac{7}{2}}\omega_{60},0,0,0,\omega_{60},0,0,0,-\sqrt{\frac{7}{2}}\omega_{60},0,0)$                                                                                               \\%\hline
    7& $(0,0,0,0,0,0,0,0,0,0,0,0,0,0,0)$                                                                                               \\%\hline
    8& $(\sqrt{\frac{65}{198}}\omega_{80},0,0,0,\sqrt{\frac{14}{99}}\omega_{80},0,0,0,\omega_{80},0,0,0,\sqrt{\frac{14}{99}}\omega_{80},0,0,0,\sqrt{\frac{65}{198}}\omega_{80})$                                                                                               \\ \hline\hline
  \end{tabular*}
  \caption{Possible values of $\omega_{lm}$ in the rest frame.}
  \label{tab:symmetryw1}
\end{table}

\section{Toy Model: Scalar QED \label{sec:toymodel}}

\subsection{The Lagrangian}

In order to test the formalism, we use scalar QED with a Higgs mechanism, since the vector state needs to be massive. This model was, for instance, studied in Ref. \cite{SCALARQED}, whose parameters will be used as a guideline. The continuum Euclidean Lagrangian of such a theory reads
\begin{equation}
 \mathcal{L}_E = \frac{1}{4} F^{\mu\nu}F_{\mu\nu} + (D_\mu \phi_c)^\dagger D_\mu \phi_c + m^2_0 |\phi_c|^2 +\lambda_c |\phi_c|^4, \label{eq:toymodel}
\end{equation}
with $D_\mu \phi_c = \partial_\mu \phi_c +i g A_\mu \phi_c$ and $F_{\mu\nu} = \partial_\mu A_\nu - \partial_\nu A_\mu$.

For the discretization we restrict space-time to a discrete and finite set of points
\[
x \in \{(x_0, x_1, x_2, x_3)\ |\ x_0 = 0, 1, ..., T,\ x_i = 0, 1, \ldots, L-1,\ i=1,2,3\}\,.
\]
We use periodic boundary conditions.
In order to include the gauge symmetry in the discretized action, one defines the discretized gauge links at point $x$ in direction $\mu$ as
\begin{equation}
  \label{eq:glink}
  U_{x,\mu}=e^{i a g A_{x,\mu}} 
\end{equation}
with gauge potential $A_{x,\mu}$, gauge coupling $g$ and lattice spacing $a$.
In the case of QED $U_{x,\mu}\in U(1)$.
Scalar field $\phi$, covariant derivative and integrals are discretized as follows:
\begin{eqnarray}
 a\phi_c(x) &\rightarrow& \sqrt{\kappa} \phi_x, \nonumber\\[2mm]
 D_\mu \phi_c(x) &\rightarrow& \kappa(U_{x,\mu}\phi_{x+\mu} - \phi_x)/a^2, \nonumber\\[2mm]
 \int d^4 x &\rightarrow& a^4 \sum_x\,.
\end{eqnarray}
This way, the discretized action reads
\begin{eqnarray}
 S = \sum_x \Big(-\frac{\beta}{2}\sum_{\mu<\nu}(U_{x,\mu\nu} + U_{x,\mu\nu}^*) -\kappa \sum_\mu(\phi_x^* U_{x,\mu}\phi_{x+\mu} +cc) + \lambda(|\phi_x|^2 - 1)^2 +  | \phi_x|^2  \Big),
\end{eqnarray}
with
\begin{eqnarray}
 \lambda_c = \frac{\lambda}{\kappa^2}, \quad (a m_{0})^2 =   \frac{1-2\lambda-8\kappa}{\kappa}, \quad \beta = \frac{1}{g^2}\,.
\end{eqnarray}
The plaquette at point $x$ in the $\mu$-$\nu$ plane is defined as usual by the smallest closed loop 
\begin{equation}
 U_{x,\mu\nu} = U_{x,\mu} U_{x+\mu,\nu} U_{x,\nu}^\dagger U_{x+\nu,\mu}^\dagger\,. 
\end{equation}

\subsection{Construction of the operators}

Any transformation of the group $\mathcal{G}$, acting on the components of any vector, is a combination of an interchange of its components, an inversion of an axis and an inversion of all axes.
%The transformations of the group $\mathcal{G}$, acting on the components of any vector, lead to:
%\begin{itemize}
%\item interchange of its components;
%\item inversion of one of the axis;
%\item inversion of all axis.
%\end{itemize}
We would like to study the transformation properties of the operators
\begin{equation}
\mathcal{O}_i(x) =  \phi^\dagger_{x} U_{x,i} \phi_{x + i} \label{eq:operatorbasic}
\end{equation}
with respect to the transformations from the group $\mathcal{G}$ (here, we choose
the spatial component $\mu=i$ of the link $U_{x,\mu}$). We will consider everything in the continuum first and then its equivalent for the discretized model.

The transformation of the scalar fields reads
\begin{equation}
\phi(x)\to\phi(x')\, ,\quad\quad x_i'=T_{ij}(\mathcal{S}^{-1})x_j\, ,\quad t'=t\, ,
\end{equation}
where the matrices $T_{ij}(\mathcal{S}^{-1})$ form a three-dimensional irrep of the cubic group
in the Cartesian basis.  Next, we consider the transformation of the link. The transformation law for the vector field is given by
\begin{equation}
A_i(x)\to T_{ij}(\mathcal{S}^{-1})A_j(x')\, .
\end{equation}
For the transformation of a link under $\mathcal{G}$ multiple cases have to be taken into account:
\begin{itemize}
\item The interchange of the components does not 
affect the index $i$. For example, $i=1$, whereas the components $2,3$ are 
interchanged. Then, in the continuum, the link transforms as
\begin{eqnarray}
U(x,x+ae_1)\to\exp\biggl(ig\int_0^1d\tau aA_1(x_1+a\tau,x_3,x_2,t)\biggr)
=U(x',x'+ae_1)\, ,
\end{eqnarray}
where $e_i$ denotes a unit vector in the direction $i$ and $x'=(x_1,x_3,x_2,t)$. 
On the lattice, this corresponds
to
\begin{equation}
U_{x,i}\to U_{x',i}\, .
\end{equation}

\item
The interchange involves the component $i$, e.g., $i=1$
and the components $1,2$ are interchanged. Then,
\begin{eqnarray}
U(x,x+ae_1)\to\exp\biggl(ig\int_0^1d\tau aA_2(x_2,x_1+a\tau,x_3,t)\biggr)
=U(x',x'+ae_2)\, ,
\end{eqnarray}
or, on the lattice,
\begin{eqnarray}
U_{x,i}\to U_{x',j}\, ,
\end{eqnarray}
where $x'$ is obtained from $x$ by interchanging the components $x_i$ and $x_j$.

Both transformations can be written as
\begin{eqnarray}
U_{x,i}\to T_{ij}(\mathcal{S}^{-1})U_{x',j}\, ,\quad\quad 
x'_i=T_{ij}(\mathcal{S}^{-1})x_j\, .
\end{eqnarray}
In other words, the link $U_{x,i}$ behaves like a vector under such transformations,
albeit not being a vector with respect to the rotation group.

\item
$i=1$ and the reflection of all axes. The result is given by
\begin{eqnarray}
U(x,x+ae_1)\to\exp\biggl(-ig\int_0^1d\tau aA_1(-x_1-a\tau,-x_2,-x_3,t)\biggr)
=U^\dagger(x'-ae_1,x')\, .
\end{eqnarray}
or, on the lattice,
\begin{eqnarray}
U_{x,i}\to U^\dagger_{x'-i,i}\, .
\end{eqnarray}

\item The inversion of one of the axes. Here, again, one has to consider
two different possibilities. First, if the axis $i$ is not affected by inversion, then
\begin{eqnarray}
U_{x,i}\to U_{x',i}\,,
\end{eqnarray}
otherwise
\begin{eqnarray}
U_{x,i}\to U^\dagger_{x'-i,i}\, .
\end{eqnarray} 
\end{itemize}

Finally, let us consider the set of the operators $\mathcal{O}_i(x)$, defined in Eq.~(\ref{eq:operatorbasic}) and construct the operators
\begin{eqnarray}
\bar{\mathcal{O}}_i(x)=\phi^\dagger_xU^\dagger_{x-i,i}\phi_{x-i}
\end{eqnarray}
Using the transformation properties of the scalar field and a link
it is straightforward to check that the following operator
\begin{eqnarray}\label{eq:A1}
S(x)=\sum_i(\mathcal{O}_i(x)+\bar{\mathcal{O}}_i(x)) \label{eq:scalarop}
\end{eqnarray}
transforms as $S(x)\to S(x')$ both under rotations and 
inversions. One may use this operator, for example,
 to project out the spectrum in the representation $A_1$ (rest frame).

On the other hand, it can be checked that the operator
\begin{eqnarray}\label{eq:T1}
V_i(x)=\mathcal{O}_i(x)-\bar{\mathcal{O}}_i(x) \label{eq:vectorop}
\end{eqnarray}
behaves like a vector both under the rotations and reflections. We shall use this
operator to construct the two-particle operators for the vector-vector scattering
in different irreps.
The conventions and the naming scheme of these irreps are listed in 
Appendix \ref{app:convention}.

The generalization for the case of a Wilson line of arbitrary length is given by 
\begin{align}
  \begin{split}
    \mathcal{O}_i(x)\to \mathcal{O}_i(x,N)&=\phi^\dagger_x\biggl(\prod_{n=0}^{N-1}U_{x+ni,i}\biggr)\phi_{x+Ni}\,, \\[2mm]
    \bar{\mathcal{O}}_i(x)\to \bar{\mathcal{O}}_i(x,N)&=\phi^\dagger_x\biggl( \prod_{n=1}^{N} U^\dagger_{x-ni,i}\biggr)\phi_{x-Ni}\,, \label{eq:operatorlength} 
  \end{split}
\end{align}
for which Eqs.~(\ref{eq:A1}) and (\ref{eq:T1}) do not change (in the following, in order
to simplify the notations, the dependence on $N$ is never displayed explicitly).
Note that such highly nonlocal operators are seen to improve the signal significantly.

\section{Operators}

A generic operator $\mathcal{O}_\alpha^\Gamma(x)$, transforming under a specific irrep $\Gamma$ of the group
$\mathcal{G}$, obeys the following transformation law
\begin{equation}
 \mathcal{O}^\Gamma_\alpha (x)\to R_{\beta \alpha} (\mathcal{S}) O^\Gamma_\beta(x').
\end{equation}
The prescription for constructing such operators is well known (see, e.g.,
Refs.~\cite{Bernard:2008ax,Gockeler:2012yj}). Consider first the case of
one-particle operators $\mathcal{O}({\bf x},t)$, whose transformation properties
(a scalar, vector, etc) are defined.
More specifically, let the action of the group element $\mathcal{S}$ on the field
$\mathcal{O}({\bf x},t)$ be represented by a linear matrix $A(\mathcal{S}^{-1})$
[unit matrix for scalars, $T_{ij}(\mathcal{S}^{-1})$ for vectors, etc.]. Then, it is possible 
to project out the component, contributing to a given irrep $\Gamma$.
In momentum space, the corresponding expression takes the form
\begin{equation}
 \mathcal{O}^\Gamma_\alpha (\mathbf p, t) = \sum_{x} e^{i\mathbf p \mathbf x} 
\sum_{\mathcal{S} \in \mathcal{G}} (R^\Gamma_{\alpha\beta}(\mathcal{S}))^* \ 
(A(\mathcal{S}^{-1}) \mathcal{O})(\mathbf x ,t ),
\end{equation}
where  the set of $R^\Gamma_{\alpha\beta}(\mathcal{S})$ forms the irrep $\Gamma$
 of the group $\mathcal{G}$ with index $\beta$ fixed.
 
The two-particle operator with total momentum $\mathbf p$ and
relative momentum $\mathbf q$ is given by 
\begin{equation}
 \mathcal{O}^\Gamma_\alpha (\mathbf p, \mathbf q, t) 
= \sum_{\mathbf x,\mathbf y} \left( \sum_{\mathcal{S} \in \mathcal{G}} e^{i\mathbf{px} + \tilde{\mathbf q}(\mathbf y-\mathbf x)} \right)  (R^\Gamma_{\alpha\beta}(\mathcal{S}))^* \ (A(\mathcal{S}^{-1})
 \mathcal{O}) (\mathbf x,\mathbf y ,t ), \label{eq:op2part}
\end{equation}
where the vector $ \tilde{\mathbf q}$ is obtained from the vector
 $ \mathbf q$ via $\tilde q_i=T_{ij}(\mathcal{S}^{-1})q_j$.

In order to simplify the construction of the operators, we note that the irreducible 
operators transform exactly as the basis vectors in the corresponding irrep. We shall illustrate the procedure with one example. Consider the construction of the two-particle
operator in the case where the momenta of the particles are $\frac{2\pi}{L}(0,0,1)$ 
and $\frac{2\pi}{L}\,(0,0,0)$.
This is a case of the little group $C_{4v}$. From Table~\ref{tab:BV2a} one finds that,
e.g., the state $\ket{2,0}$ is the basis vector in the irrep $A_1$. On the other hand,
various linear combinations of the Cartesian components of the vector field $V_i(\mathbf x,t)$ transform as
\begin{eqnarray}
   \ket{1,\pm 1}& \sim& \mp \frac{1}{\sqrt{2}}(V_1(\mathbf x,t) \pm i V_2(\mathbf x,t)), 
\nonumber\\[2mm]
 \ket{1,\pm 0}& \sim& V_3(\mathbf x,t).
 \label{eq:identif}
\end{eqnarray}
The state $\ket{2,0}$ can be obtained as a linear combination of the spin-1 states:
\begin{equation}
\ket{2,0} = \frac{1}{\sqrt{6}}\ket{1,1}\ket{1,-1} + \sqrt{\frac{2}{3}}\ket{1,0}\ket{1,0} + \frac{1}{\sqrt{6}}\ket{1,-1}\ket{1,1},
\end{equation}
Taking into account the Eq.~(\ref{eq:identif}), we finally obtain that the following
operator
\begin{equation}
\mathcal{O}^{A_1}(\mathbf p, t) 
= \sum_{\mathbf x,\mathbf y} e^{i\mathbf{px}} (-V_1(\mathbf x,t)V_1(\mathbf y,t)-V_2(\mathbf x,t)V_2(\mathbf y,t)+2V_3(\mathbf x,t)V_3(\mathbf y,t))\, ,
\end{equation}
with $\mathbf p=\frac{2\pi}{L}\,(0,0,1)$, indeed projects on the irrep $A_1$ of the group
$C_{4v}$.

We have collected one- and two-particle operators in 
Table \ref{tab:op1particle} and Table \ref{tab:op2particle},
respectively. Note that this simplified procedure is only possible if one of the momenta is zero or both are in the same little group;  if that is not the case, one must use Eq.~(\ref{eq:op2part}), as in the case for $\Gamma=A_1$, $\mathbf{p}=\frac{2\pi}{L}(1,1,0)$ and $\mathbf{q}=\frac{2\pi}{L}(1,1,0)$ of Table \ref{tab:op2particle}.

\begin{minipage}[b]{1\linewidth}
\begin{table}[H]
\centering
\begin{tabular}{| c | c | c |}
\hline
$\mathbf d$  & $\Gamma$ & Operator \\ \hline
$(0,0,0)$ & $T^-_1$  & $V_i({\mathbf x},t) $ \\ \hline 
\multirow{3}{*}{$(0,0,1)$} & $A_1$  & $V_3({\mathbf x},t) $ \\ \cline{2-3}
  & \multirow{2}{*}{$E$}  & $V_1({\mathbf x},t) + V_2({\mathbf x},t)$ \\ \cline{3-3}
  &   & $V_1({\mathbf x},t) - V_2({\mathbf x},t)$ \\ \hline
\multirow{3}{*}{$(1,1,0)$} & $A_1$  & $ V_1({\mathbf x},t)+V_2({\mathbf x},t) $ \\ \cline{2-3}
  & $B_1$ & $V_1({\mathbf x},t) - V_2({\mathbf x},t)$ \\ \cline{2-3}
  & $B_2$ & $V_3({\mathbf x},t)$ \\ \hline
\multirow{3}{*}{$(1,1,1)$} & $A_1$  & $\sum_{i=1}^3 V_i({\mathbf x},t)$ \\ \cline{2-3}
  & \multirow{2}{*}{$E$} & $V_1({\mathbf x},t) - V_2({\mathbf x},t)$ \\ \cline{3-3} 
  &  & $V_1({\mathbf x},t) +V_2({\mathbf x},t) - 2V_3({\mathbf x},t)$ \\ \hline 
\end{tabular}  
\caption{Complete list of one-particle operators in the multiple moving frames and irreps. \label{tab:op1particle}}
\end{table}
\end{minipage}
\begin{center}
  \begin{table}[H]
%   \begin{table}
    \centering
    \begin{tabular*}{.7\textwidth}{@{\extracolsep{\fill}}ccccccc}
      \hline\hline
      $L^3 \times T$  & Ref. & N    &$m^2_0$ & $\lambda_c$ & $\kappa$ & $\lambda$ \\
      \hline\hline
      $16^3 \times 32$&   A16   & 138000 &-35     & 88          & 0.18425 & 2.9873    \\
      \hline
      \multirow{5}{*}{$12^3 \times 24$}&   A12   & 33000 &-35     & 88          & 0.18425 & 2.9873    \\
      &   B12   & 24000 &-35.5     & 90.6          & 0.18208 & 3.0036    \\ 
      &   C12   & 21500 &-35.6     & 91.6          & 0.18084 & 2.9956    \\ 
      &   D12   & 16000 &-35.85    & 93.1          & 0.17949 & 2.9994    \\ 
      &   E12   & 18800 &-36.1     & 94.7          & 0.17802 & 3.0012    \\ \hline\hline
    \end{tabular*}  
    \caption{Ensembles used for the simulations. The gauge coupling is kept constant, $\beta = 2.5$. \label{tab:ensembles12}}
  \end{table}
\end{center}
\begin{minipage}[b]{1\linewidth}
\begin{table}[H]
\centering
\begin{tabular}{| c | c | c | c | c |}
\hline
$\mathbf d$  & $\Gamma$ & $\frac{L}{2\pi} \mathbf p$ & $\frac{L}{2\pi} \mathbf q$ & $\mathcal{O}^\Gamma (\mathbf x, \mathbf y,t)$ \\ \hline
\multirow{3}{*}{$(0,0,0)$} & $A_1$  & $\mathbf{0}$ & $\mathbf{0}$ & $\sum_{i=1}^3 {V}_i({\mathbf x},t){V}_i({\mathbf y},t)$\\ \cline{2-5}
  & $E^+$  & $\mathbf{0}$ & $\mathbf{0}$ & $V_1({\mathbf x},t){V}_1({\mathbf y},t) - {V}_2({\mathbf x},t){V}_2({\mathbf y},t)$\\ \cline{2-5}
  & $T_2^+$  & $\mathbf{0}$ & $\mathbf{0}$ & 
${V}_1({\mathbf x},t){V}_2({\mathbf y},t)+{V}_2({\mathbf x},t){V}_1({\mathbf y},t)$  \\ \hline 
\multirow{5}{*}{$(0,0,1)$} & $A_1$   & $(0,0,n)$ &$(0,0,m)$ & $V_1(\mathbf x,t)V_1(\mathbf y,t)+V_2(\mathbf x,t)V_2(\mathbf y,t)+V_3(\mathbf x,t)V_3(\mathbf y,t))$  \\ \cline{2-5}
 & $A_1$   & $(0,0,n)$ &$(0,0,m)$ & $-V_1(\mathbf x,t)V_1(\mathbf y,t)-V_2(\mathbf x,t)V_2(\mathbf y,t)+2V_3(\mathbf x,t)V_3(\mathbf y,t))$  \\ \cline{2-5}
  & $A_2$   & $(0,0,n)$ & $(0,0,m)$ & $V_1(\mathbf x,t)V_2(\mathbf y,t)-V_2(\mathbf x,t)V_1(\mathbf y,t)$ \\ \cline{2-5}
  & $B_1$  & $(0,0,n)$ & $(0,0,m)$ & $V_1(\mathbf x,t)V_1(\mathbf y,t)-V_2(\mathbf x,t)V_2(\mathbf y,t)$   \\ \cline{2-5}
  & $B_2$  & $(0,0,n)$ & $(0,0,m)$ & $V_1(\mathbf x,t)V_2(\mathbf y,t)+V_2(\mathbf x,t)V_1(\mathbf y,t)$ \\ \cline{2-5}
  & $E $  & $(0,0,n)$ & $(0,0,m)$ & $V_3(\mathbf x,t)(V_1(\mathbf y,t)+V_2(\mathbf y,t))+(V_1(\mathbf x,t)+V_2(\mathbf x,t))V_3(\mathbf y,t)$ \\ \cline{1-5}
  \multirow{3}{*}{$(1,1,0)$} & $A_1$   & $(1,1,0)$ & $\mathbf 0$ & \multirow{2}{*}{$V_3(\mathbf x,t)V_3(\mathbf y,t)$}  \\ \cline{2-4}
  & $A_1$   & $(1,1,0)$ & $(0,1,0)$ &   \\ \cline{2-5}
  & $A_2$   & $(1,1,0)$ & $\mathbf 0$ & $V_3(\mathbf x,t)(V_1(\mathbf y,t)-V_2(\mathbf y,t))+(V_1(\mathbf x,t)-V_2(\mathbf x,t))V_3(\mathbf y,t)$  \\ \cline{1-5}
 
%\multirow{3}{*}{$(1,1,0)$} & $A_1$  & $ V_1(x)+V_2(x) $ \\ \cline{2-3}
%  & $B_1$ & $V_1(x) - V_2(x)$ \\ \cline{2-3}
%  & $B_2$ & $V_3(x)$ \\ \hline
%\multirow{3}{*}{$(1,1,1)$} & $A_1$  & $\sum_{i=1}^3 V_i(x)$ \\ \cline{2-3}
%  & \multirow{2}{*}{$E$} & $V_1(x) - V_2(x)$ \\ \cline{3-3}
%  &  & $V_2(x) - V_3(x)$ \\ \hline
\end{tabular}  
\caption{List of used two-particle operators in the multiple moving frames and irreducible representations. The operator $\mathcal{O}^\Gamma (\mathbf p, \mathbf q,t)$ is built from the position-space operators, given in this table, by calculating the Fourier-transform with $ e^{ i\mathbf{px}+i\mathbf q(\mathbf y-\mathbf x)}$. This prescription holds for all operators except with $\mathbf p=\frac{2\pi}{L}(1,1,0)$ and $\mathbf q= \frac{2\pi}{L}(0,1,0)$ (the second line from below), for which the shortcut is no more applicable and one has to use Eq.~(\ref{eq:op2part}). \label{tab:op2particle}}
% \blue{ The complete operator $\mathcal{O}^\Gamma (\mathbf p, \mathbf q,t)$ is to be built using the conventions of Eq.~(\ref{eq:op2part}), in particular with the correct factor for the fourier transformation. }
%\red{There was probably a typo in the seond line from the last. However, taking into 
%account the statement ``Note
%that this procedure is only possible if one of the momenta is zero or both are in the same little
%group,'' it would be safer to remove this line altogether. Did one use this to perform calculations? What do you think about this?}
\end{table}
\end{minipage}

\section{Numerical Results}

The parameter sets we use are compiled in Table~\ref{tab:ensembles12}.
Note that we have five different sets of bare parameters $\lambda$ and $\kappa$ for $L=12$ and $T=24$.
For one of these parameter sets we have a second volume with $L=16$ and $T=32$.
We compute correlation functions
\begin{equation}
  C^\Gamma(t-t')\ =\ \langle \mathcal{O}^\Gamma(t) (\mathcal{O}^\Gamma)^\dagger(t')\rangle
\end{equation}
using the operators defined in the previous sections.
At large time differences $t-t'$ these correlation functions are proportional to $\exp(-E(t-t'))$ with $E$ the energy of the lowest state with the corresponding quantum numbers.
The energies are calculated with a fit to the shifted correlation function
\[
\tilde{C}^\Gamma(t) = C^\Gamma(t) - C^\Gamma(t+1)
\]
including an excited state, and the errors are calculated using the
Jackknife method. 
We use the shifted correlation function to subtract any contribution
constant in time stemming from vacuum expectation values, see also Ref.~\cite{Ottnad:2017bjt}.
Thermal contaminations in the two-particle
correlation functions with nonzero total momentum turn out to be not
important for our analysis (see plots in Appendix \ref{app:meff}). Therefore, we
have used the shifted correlation function throughout.
%\blue{Thermal contaminations in the two particle correlators with total momentum $\mathbf{P} \neq 0$ seem not to be important (See plots in Appendix \ref{app:meff}) and they have not been particularly treated}.
All the results are listed in the tables of Appendix \ref{app:results} and they will be discussed in this section.
In addition, we show exemplary plots for effective masses for selected
correlation functions in Appendix \ref{app:meff}.
The effective mass $m_\mathrm{eff}$ is calculated by solving
\begin{equation}
\frac{C^\Gamma(t) - C^\Gamma(t+1)}{C^\Gamma(t+1) - C^\Gamma(t+2)} = \frac{{\sinh(m_\mathrm{eff}(t-T/2+a/2))}}{\sinh(m_\mathrm{eff}(t+1-T/2+a/2))}
\end{equation}
numerically for $m_\mathrm{eff}$.

\subsection{One-particle results}

\begin{figure}
   \centering
   \subfigure[Mass of the vector particle for ensemble A12 for different lengths of the operator in Eq.~(\ref{eq:vectorop}). \label{fig:imL}]%
            {\includegraphics[width=0.475\textwidth,clip]{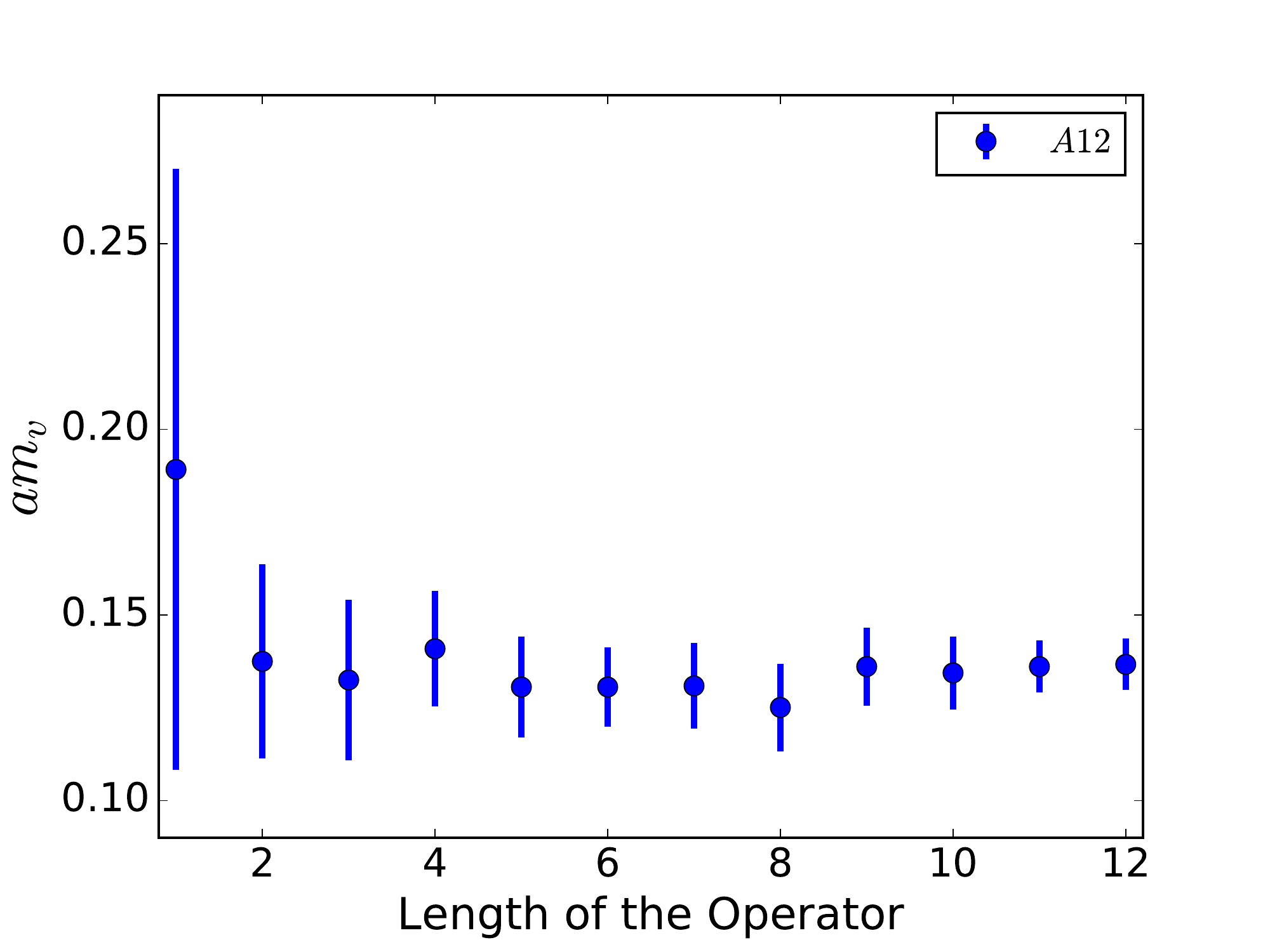}}\hfill   
              \subfigure[Mass of the scalar and vector particle for $L=12$ as a  function of $\kappa$. \label{fig:spectrum}]%
             {\includegraphics[width=0.475\textwidth,clip]{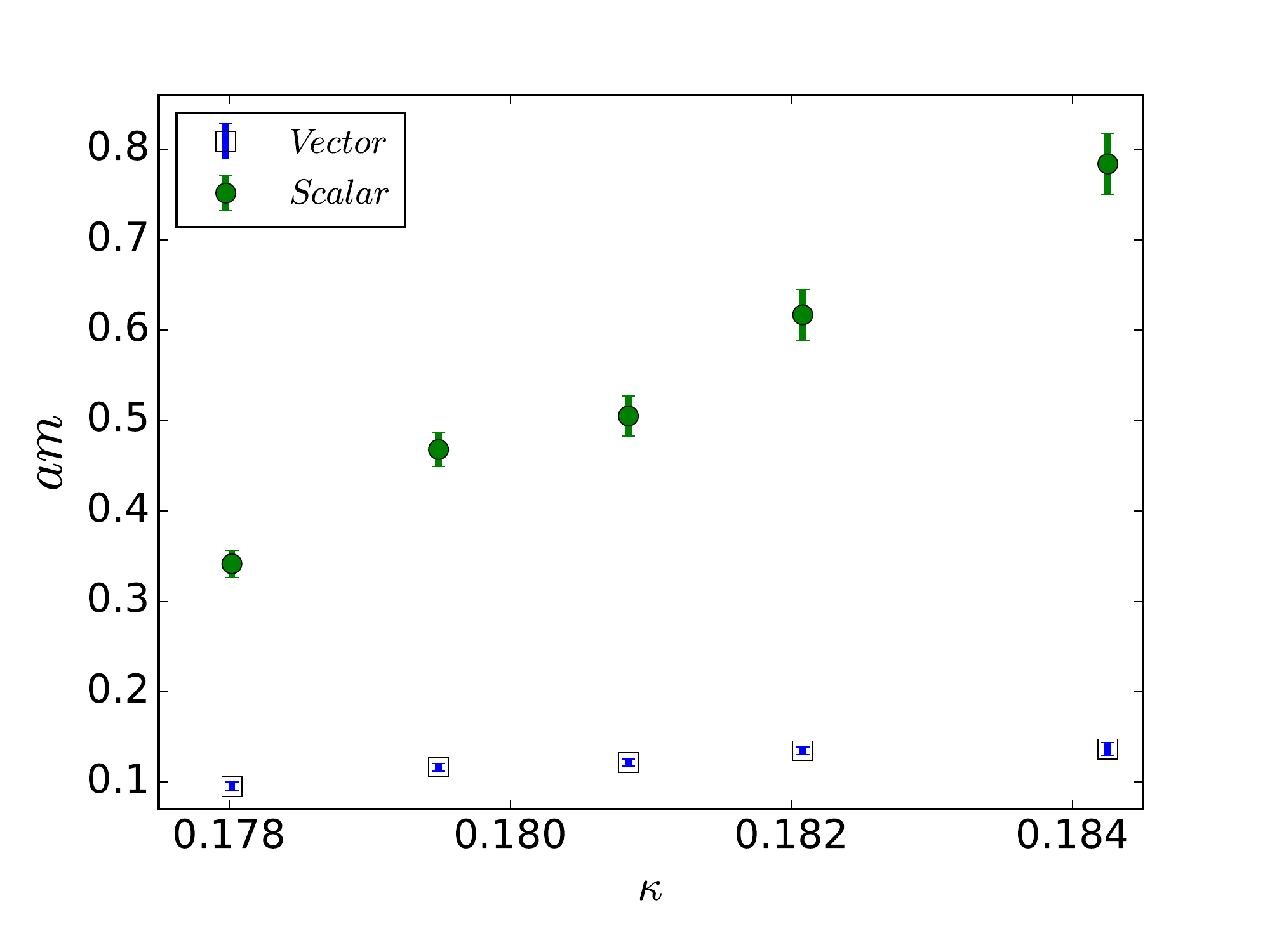}}\hfill
   \subfigure[Mass of the scalar particle for ensemble A12 as a function of the length of the operator, $N$, as in Eq.~(\ref{eq:operatorlength}). {Further explanation can be found in the text. Note that the error  in $m_{VV}^{A_1}$ is too small to be seen.}
\label{fig:reL}]%
             {\includegraphics[width=0.475\textwidth,clip]{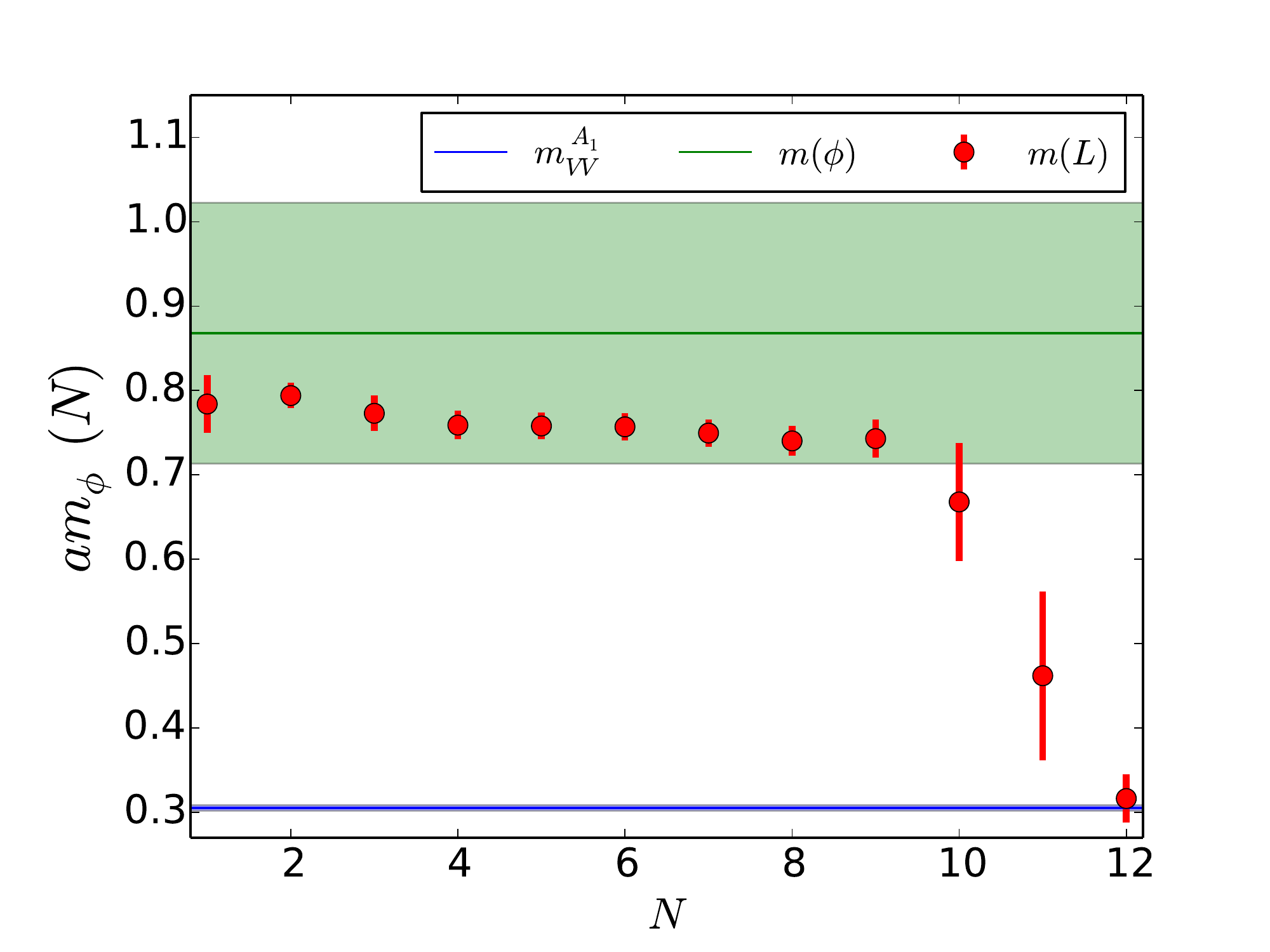}}  
   \subfigure[{Energy of the vector particle for different irreps
       in the first moving frame for $L=12$ as a function of
       $\kappa$. It is also compared with the rest frame result by
       means of the continuum dispersion relation. \label{fig:moving12}}]%
             {\includegraphics[width=0.475\textwidth,clip]{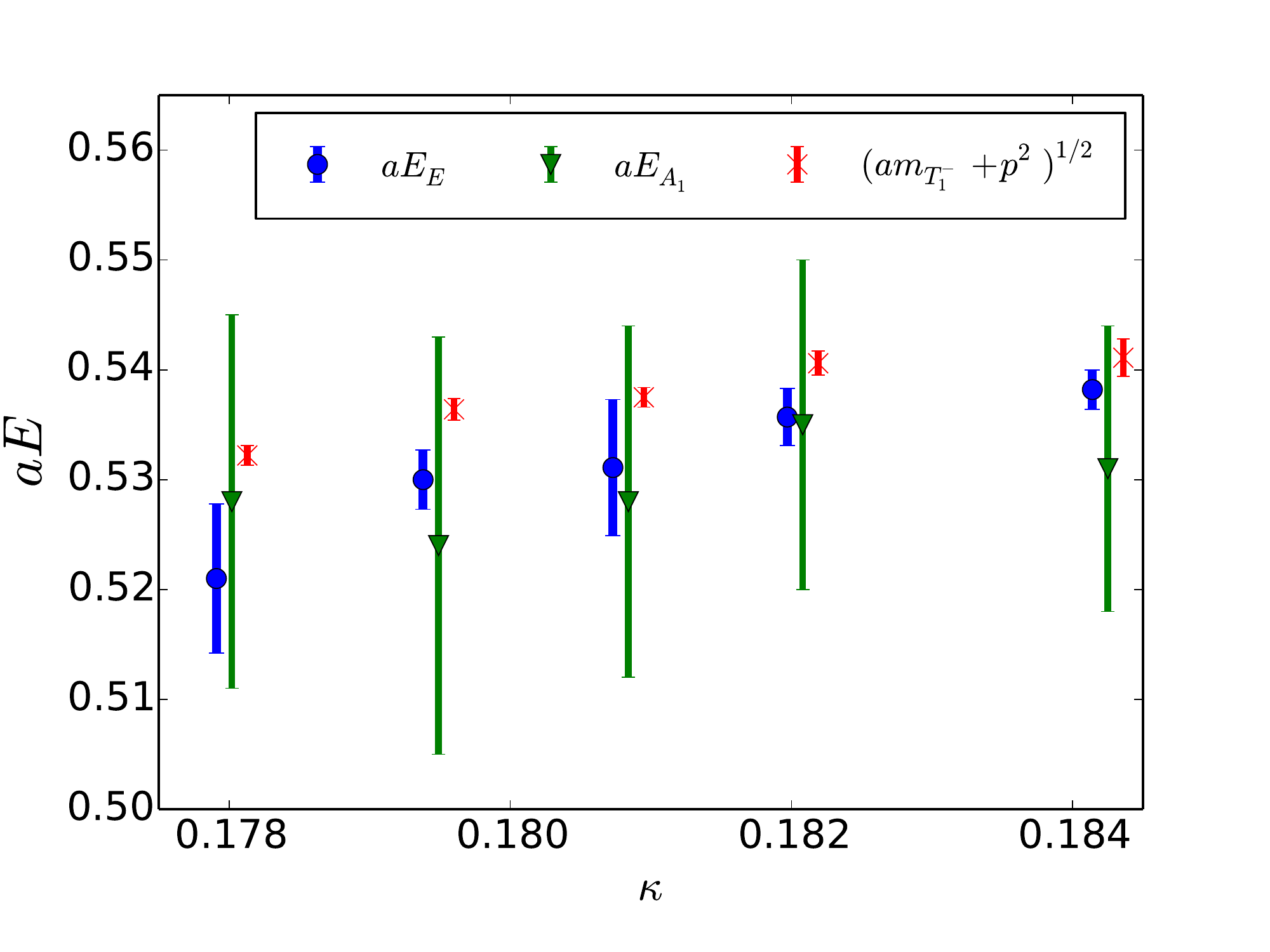}}
%\end{figure}
%\begin{figure}\ContinuedFloat             
            \caption{Numerical results for single particle operators in ensembles with L=12.}
   \label{fig:onepart}
\end{figure}

In Figure~\ref{fig:imL} we show the dependence of the mass of the vector
particle $m_v$ on the length of the operator in Eq.~(\ref{eq:operatorlength}) for
ensemble A12.
We observe a clear improvement of the signal with increasing operator length.
When using moving frames, the best signal is empirically seen at $N = L/(d+1)$, with $d$ being the units of momentum in that particular direction.  

Similarly, in Figure~\ref{fig:reL} we show the dependence of the mass of the scalar particle $m_s$ on the length of the operator in Eq.~(\ref{eq:scalarop}), N, for ensemble A12.
%\textbf{CU: do we want to use $m(L)$?}\red{or $m(N)$, because $L$ is the size of the box?}
The same mass can be measured using operator $\mathcal{O}=|\phi|$.
In addition we show the mass measured using the operator $\mathcal{O}^{A_1}(0,0,t)$.
The corresponding results are shown in the same figure as horizontal lines denoted as $m(\phi)$ for the operator $\mathcal{O}=|\phi|$ and $m_W^{A_1}$ for operator $\mathcal{O}^{A_1}(0,0,t)$, respectively.
For $m_s$ the signal improves again with increasing operator length.
However, while $m_s$ agrees with $m(\phi)$ up to operator lengths of nine, at lengths larger than nine its values drops and finally agrees with $m_W^{A_1}$. 
{It is expected that these two states mix because they have the same quantum numbers. Hence, it seems that the overlap of the operator Eq.~(\ref{eq:scalarop}) shifts with increasing operator length due to the presence of more gauge links. Moreover, since the mass of the scalar is only needed as a
reference, and the gap between the scalar and the vector mass is big,
we do not perform a variational analysis at this point.}
% \red{What can be concluded from this?}

In Figure~\ref{fig:spectrum} we compare the masses of a single vector and scalar particle as a function of $\kappa$ for $L=12$.
With increasing $\kappa$-value we observe the vector mass to be approximately constant while the scalar mass increases almost linearly.
In the range of $\kappa$-values studied here the vector mass value is always smaller than the scalar mass value.
We recall that in the continuum the bare masses of the particles are given by $m^2_\phi = -2m_0^2$ and $m_V^2 = -\frac{g^2m_0^2}{\lambda_c}$, respectively.
Hence, we expect the mass of the vector to be suppressed with respect to the scalar mass by a factor $g$ and $1/\lambda_c$.
However, it is not clear why the scalar mass duplicates with increasing $\kappa$, whereas the vector mass increases at best slightly.

{In Figure~\ref{fig:moving12} we show the energies of a single
  vector particle in the first moving frame as a function of $\kappa$
  for $L=12$ for irreps $E$ and $A_1$. The (red) crosses
  represent the prediction by the continuum dispersion
  relation with the rest frame mass (irrep $T_1^-$) as input.
  For the moving frames we observe larger statistical uncertainties as
  compared to the dispersion relation. The energy splittings between different
  irreps and to the dispersion relation prediction are compatible with
  zero.}

\begin{figure}[H]
   \centering
   \subfigure[{Mass of the vector particle for different irreps and total momenta for ensemble A16. The continuum dispersion relation has been used.} \label{fig:moving16}]
             {\includegraphics[width=0.48\textwidth,clip]{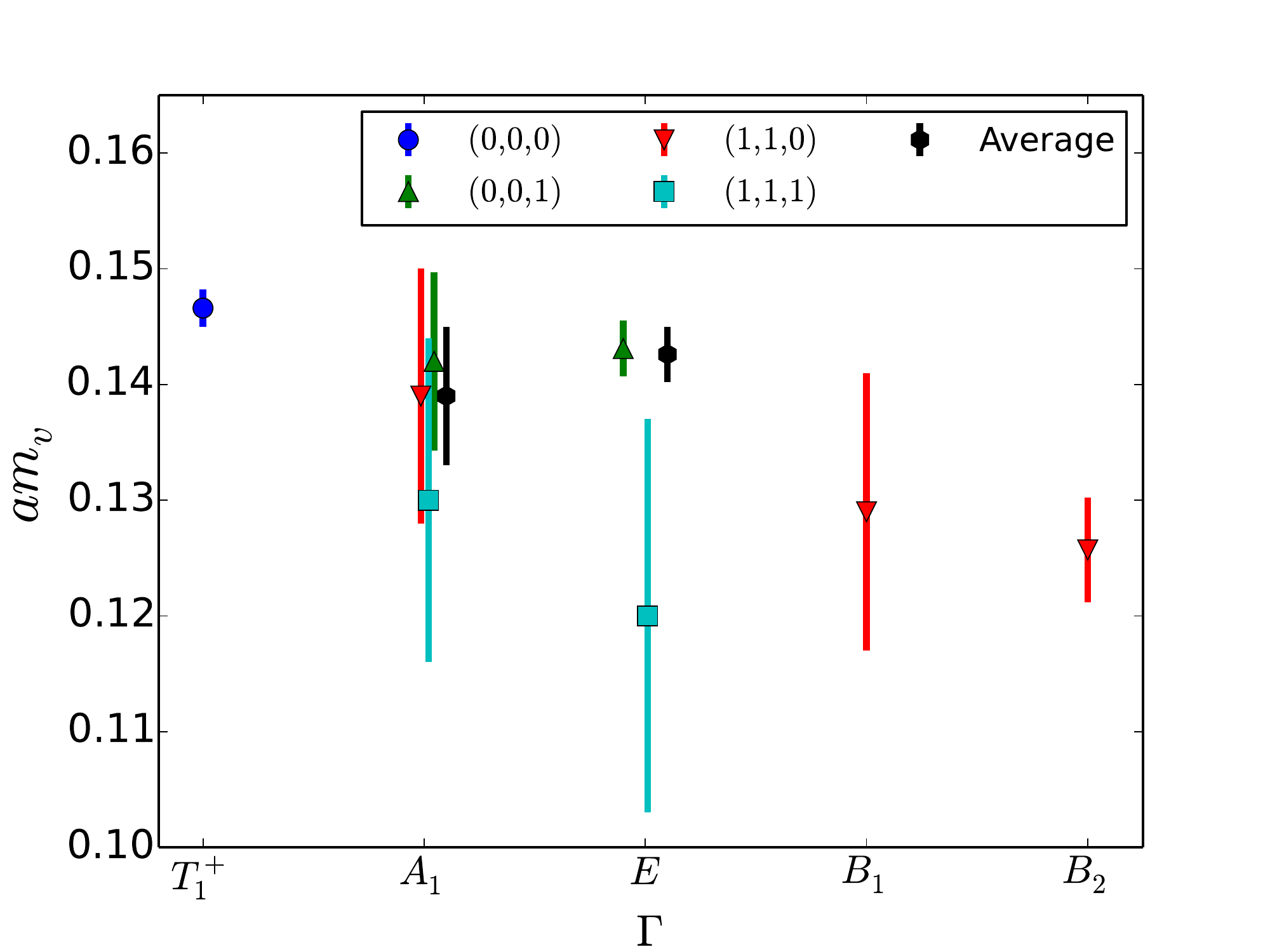}} \hfill
  \subfigure[{Energies of a single vector particle as a function of the
      squared momentum $(ap)^2$. The solid and dashed line represent
      the prediction of the continuum and lattice dispersion
      relations, respectively. Different irreps are slightly displaced
      for better readability.} \label{fig:disp}]
             {\includegraphics[width=0.48\textwidth,clip]{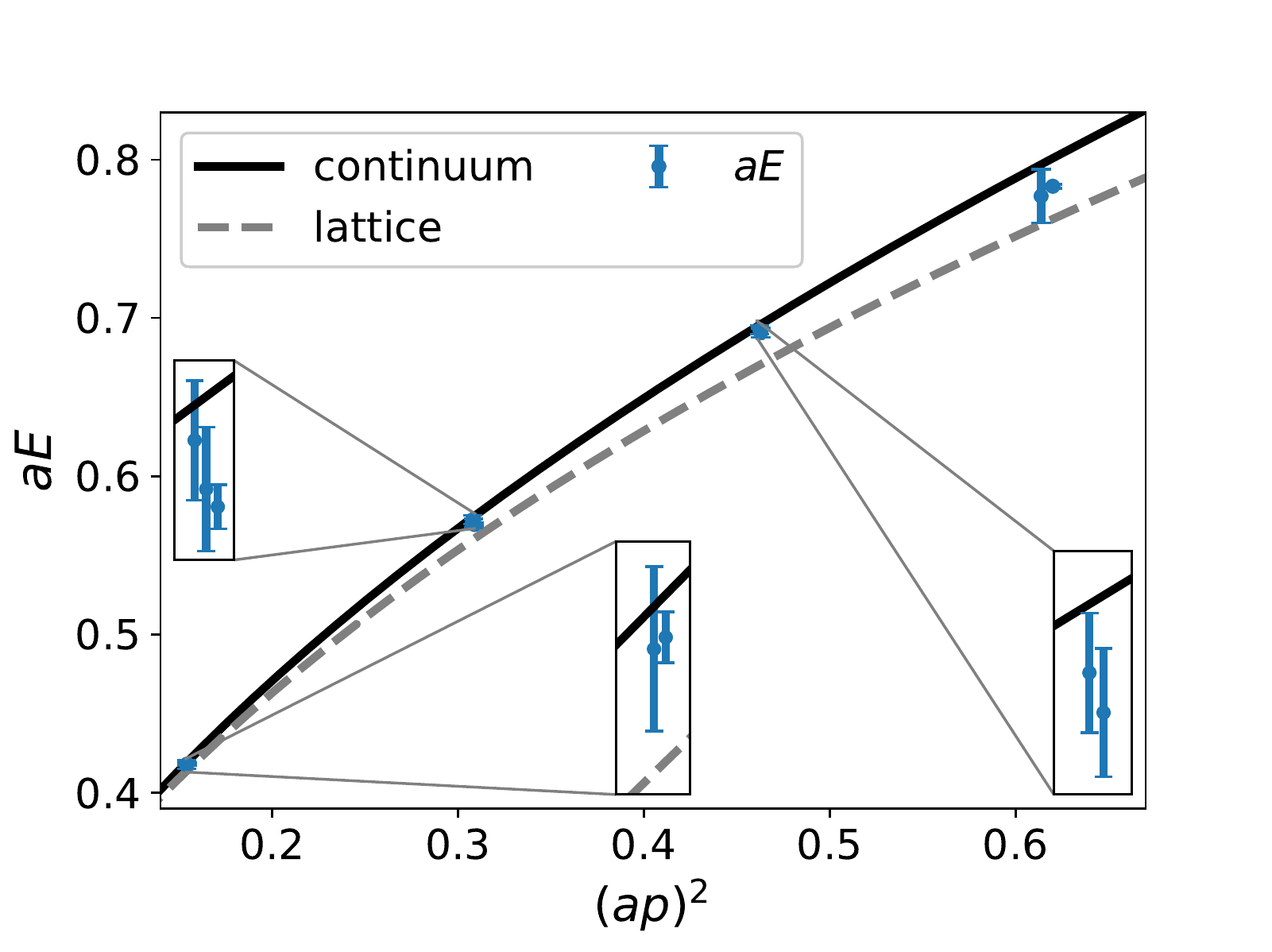}} 
   \caption{Numerical results for single particle operators in ensemble A16.}
   \label{fig:onepart2}
\end{figure}

{The comparison between different irreps is shown in more detail
  in Figure~\ref{fig:moving16}: we plot $m_v$ for ensemble A16 for
  different irreps and center of mass momenta. Where we have several
  momenta for a given irrep, we also show the weighted average value.
  The values obtained for the different moving frames tend to be smaller
  than the one in the rest frame, although in every case but one they
  are compatible within $2 \sigma$. This discrepancy may be associated
  to discretization effects.  In order to show this, in
  Figure~\ref{fig:disp} we plot for ensemble A16 the energy of the
  vector particle in different irreps and moving frames together with
  the prediction from the continuum and lattice dispersion relations,
  the latter one reading
  \begin{equation}
    \cosh aE = \cosh am + (1-\cos ap)\,.
  \end{equation}
  Here, the mass $am$ is taken to be the one measured in the rest frame, $m_V^{T^+_1}$.
  %In order to show this, we include the comparison of the continuum and lattice dispersion relation,
  %\begin{equation}
  %\cosh aE = \cosh am + (1-\cos ap),
  %\end{equation}
  %with the results for the vector particle energy in different irreps and moving frames  in Figure~\ref{fig:disp}. 
  This shows that the continuum dispersion relation describes our data
  better and we do not observe large discretization effects for $ap <
  0.6$. However,  around $ap \simeq 1$, this description becomes
  worse, as cut-off effects get bigger. From now on, we will always
  use the continuum dispersion relation with the mass as obtained from
  the rest frame as input.}

\subsection{Two-particle results}

\begin{figure}
  %\ContinuedFloat 
   \centering
   \subfigure[Energy difference $\Delta E$ as a function of $\kappa$ in the rest frame for different irreps. 
   Open symbols correspond to ensemble A16 and closed ones to  L=12.
   \label{fig:diffE}]%
             {\includegraphics[width=0.465\textwidth,clip]{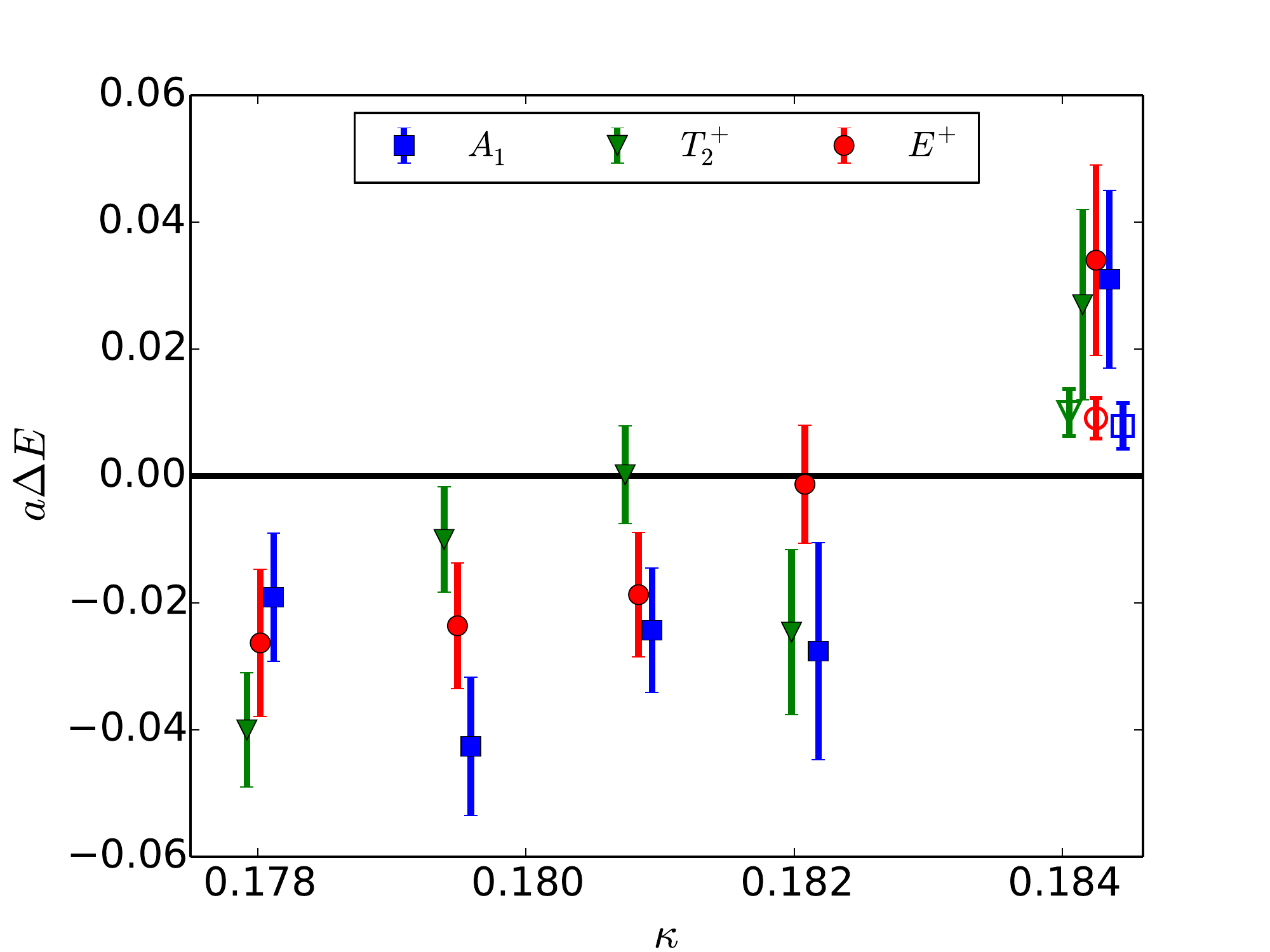}} \hfill
   \subfigure[Energy difference $\Delta E$ for ensemble A16 and A12 as a function of  the length of the box, $L$. Additionally, we include the expected behaviour.  \label{fig:diffEL}]%
             {\includegraphics[width=0.46\textwidth,clip]{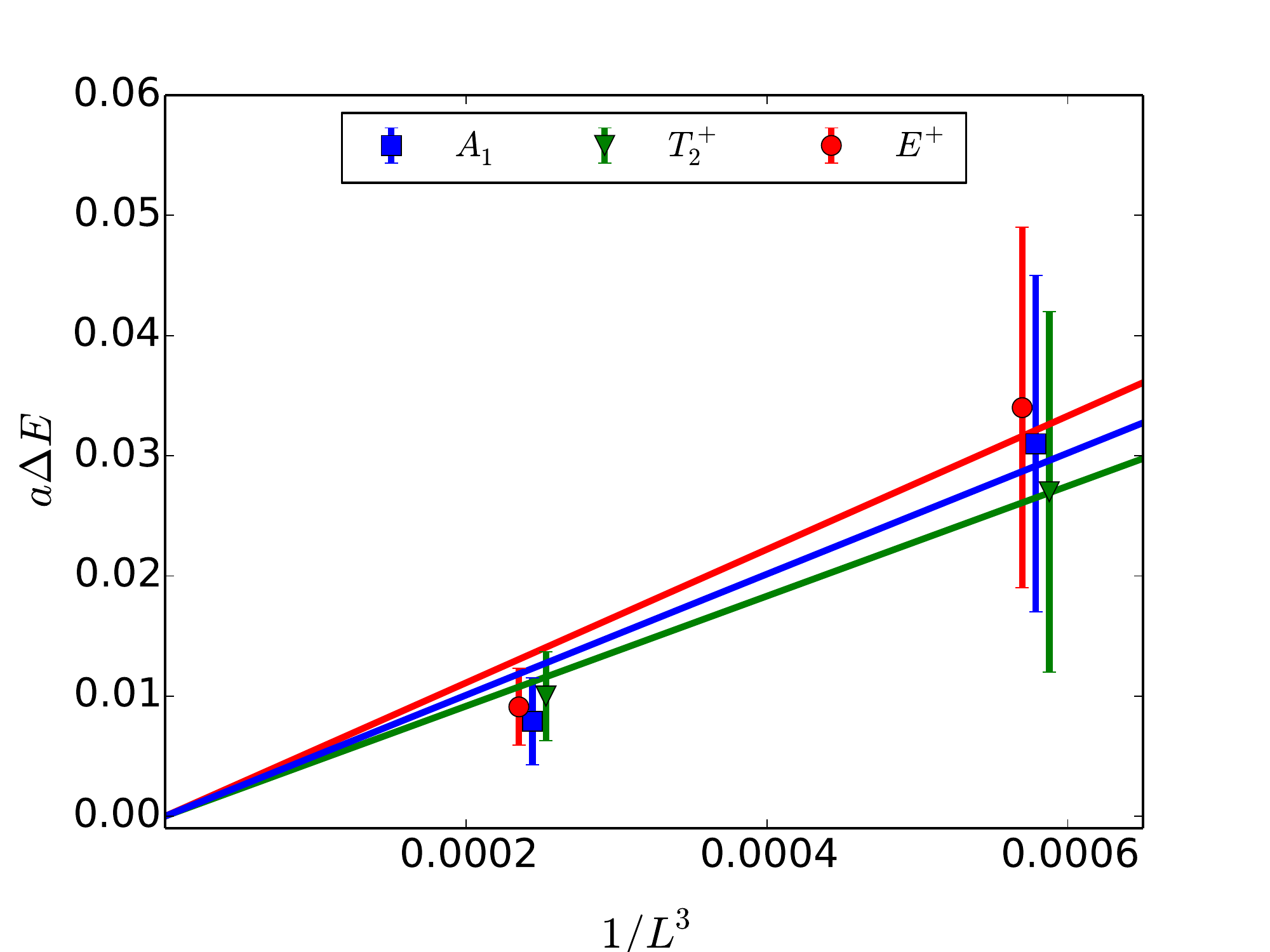}}
   \subfigure[Phase shift in the $J^P=0^+$ channel  in ensembles A12, A16 as a function of
     the scattering momentum $k$. Partial waves $J>1$ have been neglected and the two possible $L,\ S$ combinations cannot be distinguished. \label{fig:phaseshift}]%
             {\includegraphics[width=0.47\textwidth,clip]{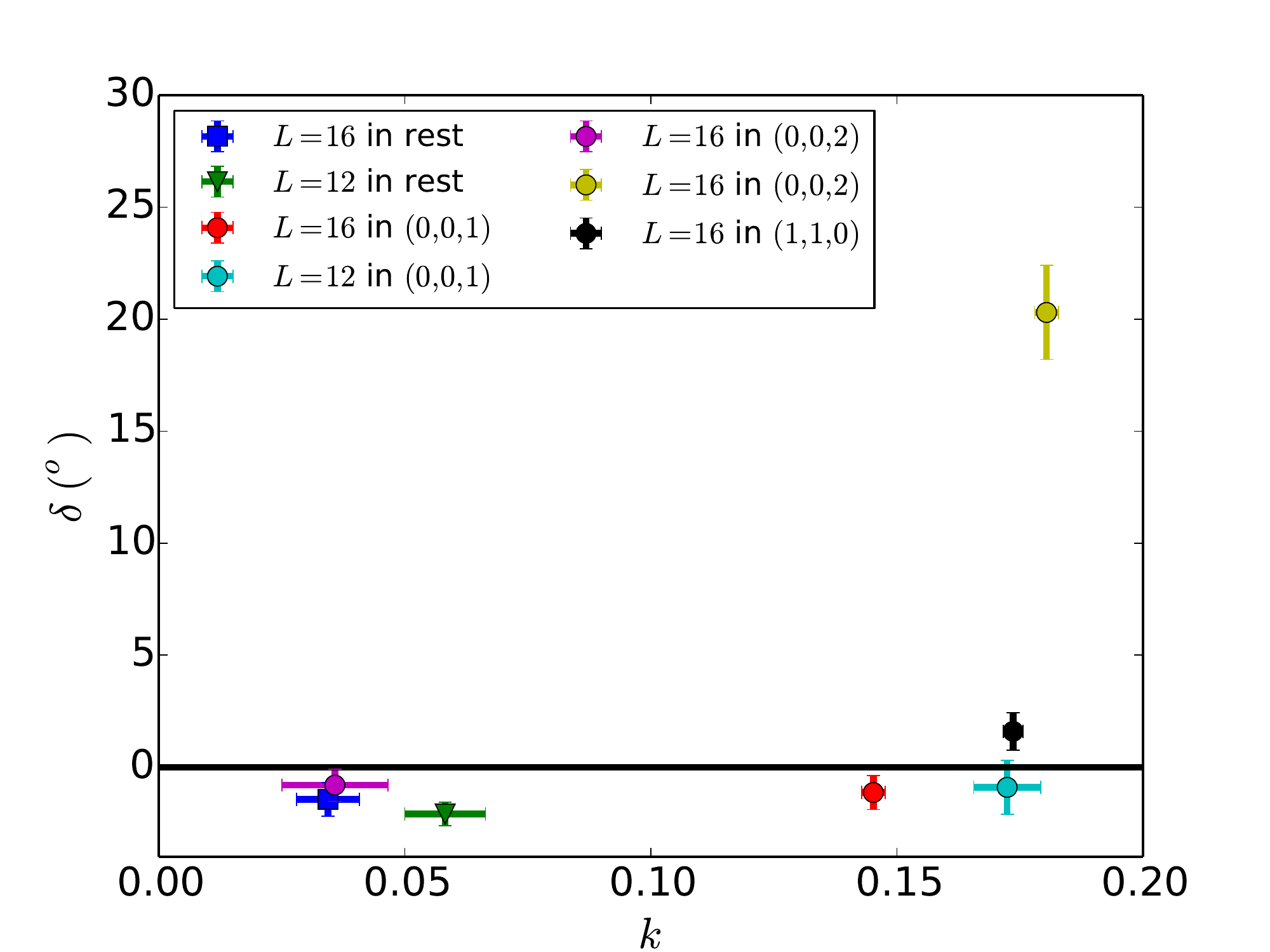}} \hfill
   \subfigure[Phase shift in the $J^P=0^-$ channel in ensemble A16 as a function of
     the scattering momentum $k$. They are calculated neglecting partial waves $J>1$. \label{fig:phaseshiftminus}]%
             {\includegraphics[width=0.47\textwidth,clip]{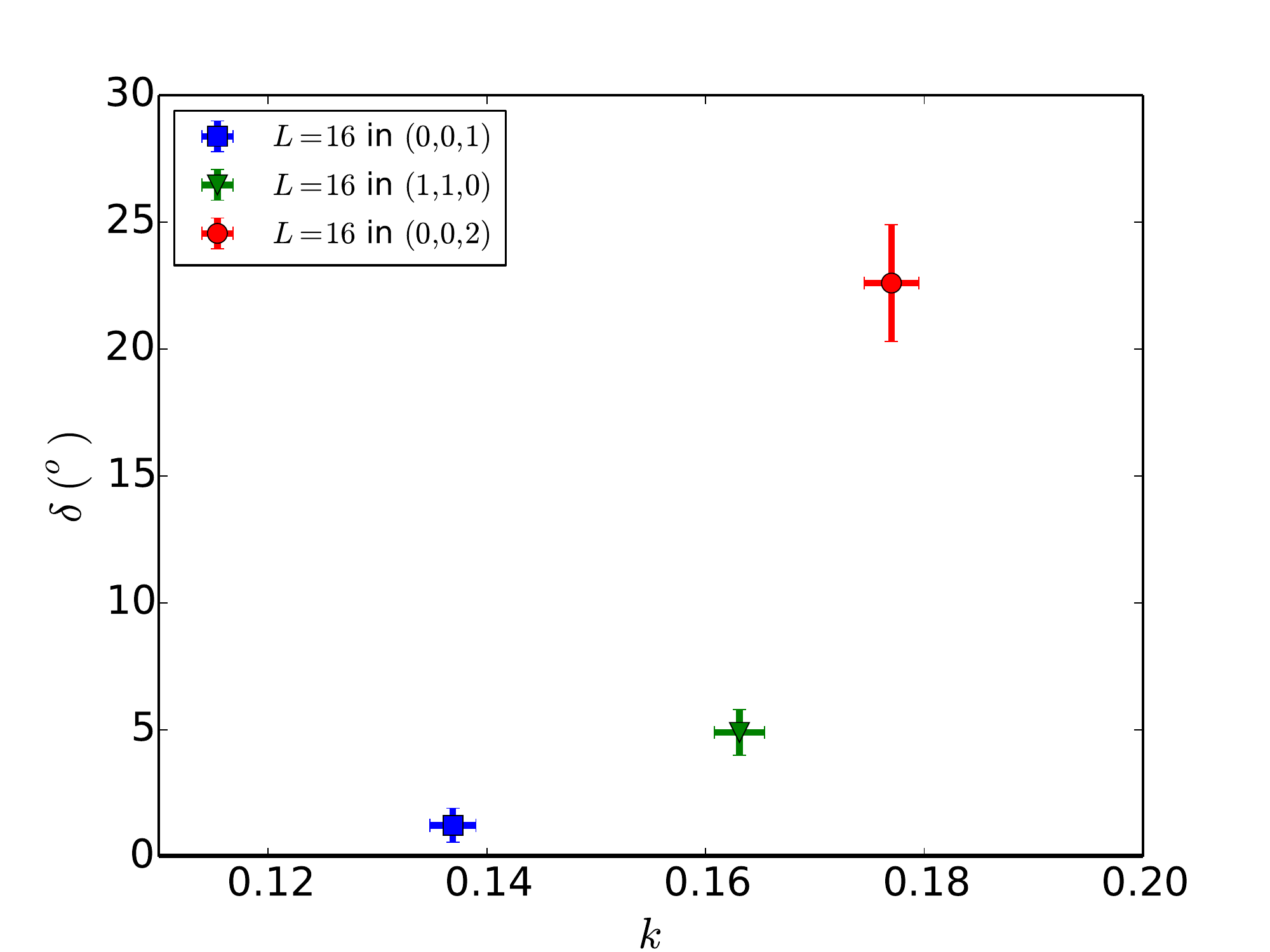}}              
   \subfigure[Energy difference $\Delta E$ for ensemble A16 in the
     moving frame with total momentum $p=\frac{2\pi}{L}(0,0,1)$ and
     $q=0$ as in Eq.~(\ref{eq:op2part}) for different irreps $\Gamma$. \label{fig:001}]%
             {\includegraphics[width=0.44\textwidth,clip]{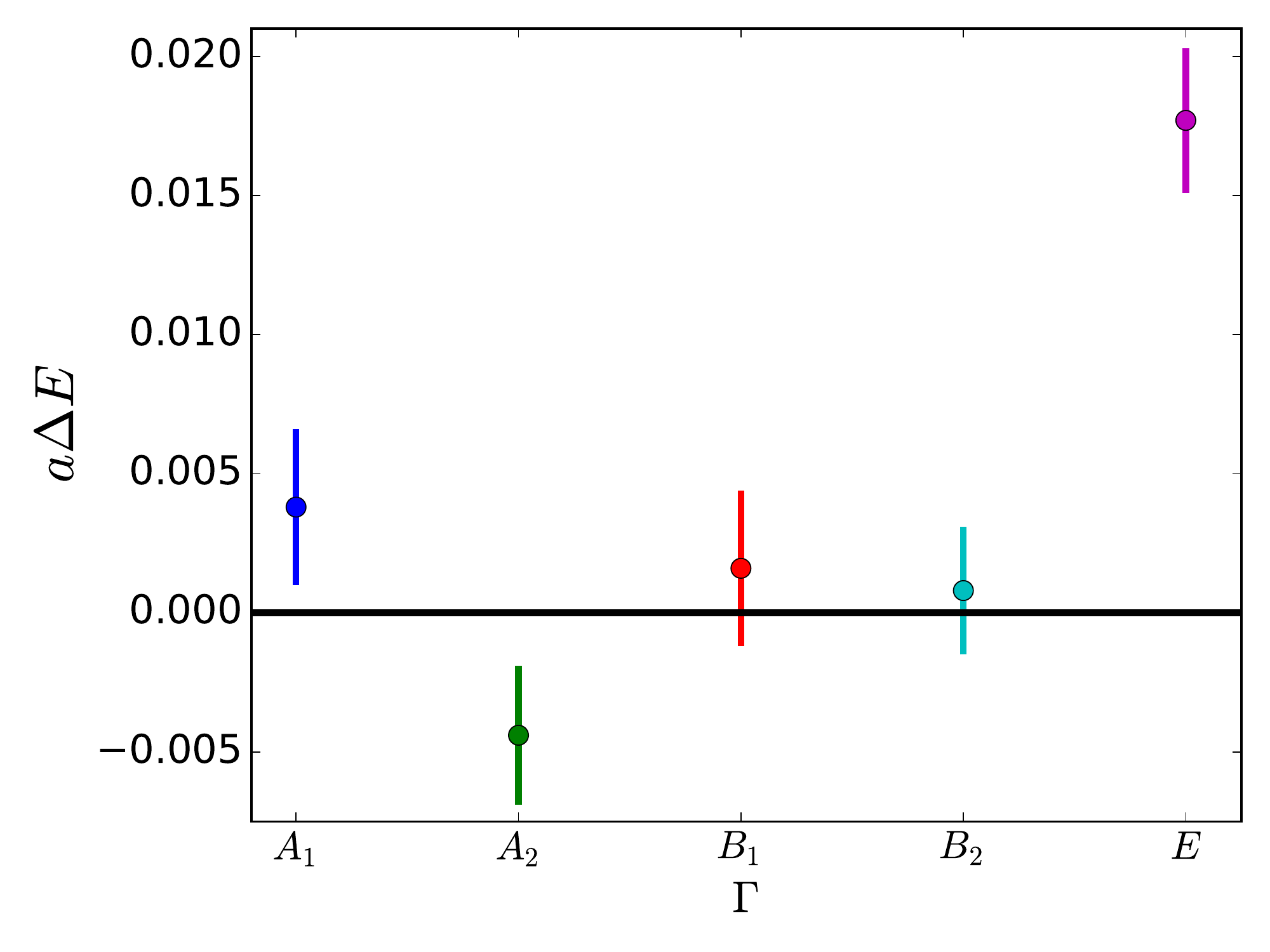}} \hfill
   \caption{Numerical results for two particles.}
    \label{fig:twopart}
\end{figure}

The energy difference $\Delta E$ is defined as the difference between
the two-particle energy on the lattice and the two-particle energy in
absence of interactions. In Figure \ref{fig:diffE} we show the results for $\Delta E$ as a function of $\kappa$ for the irreps $A_1$, $T_2^+$ and $E^+$.
For the largest $\kappa$-value we include both, A12 and A16 in the plot, respectively.
For higher $\kappa$, the interaction leads to the positive shift ($\Delta E >0$) with $\gtrapprox 2\sigma$ statistical significance.
As $\kappa$ becomes smaller, all particles become lighter and the interaction seems to flip signs.
For the lowest values of $\kappa$, the two-particle states have $\Delta E <0$.
Unfortunately, in the transition region $\Delta E$ is compatible with zero.
Comparing A12 and A16, we see the expected volume dependence in the energy shift ($\Delta E \propto L^{-3}$), when comparing $L=12$ with $L=16$.
This can be inferred from Figure~\ref{fig:diffEL}, where we show $\Delta E$ for A12 and A16 as a function of $1/L^3$.

In Figure~\ref{fig:001} we show $\Delta E$ for ensemble A16 in the first moving frame for the irreps $A_1$, $A_2$, $B_1$, $B_2$ and $E$.
As expected, $\Delta E$ depends on the considered irrep.
For the moving frame shown in Figure~\ref{fig:001} $\Delta E$ is only significantly different from zero for the $E$ irrep.

Subsequently, in Figure~\ref{fig:phaseshift} we show the phase shifts with $J^P=0^+$ computed from the energy shifts. Note that we neglect partial waves with $J>1$ and that the two possible $L,S$ combinations cannot be distinguished at this level:
\begin{equation}
\cot \delta_{0^+} = \omega_{00}.
\end{equation}
For the highest momentum shown, the ratio between the nonrelativistic kinetic energy and the mass is quite large, $\frac{E_k}{m} \approx 0.8$.
Hence, the kinematic suppression of higher partial waves, though present, is not strong anymore and a corresponding systematical error is to be expected.
The phase shift appears to be small and negative for small scattering momentum, indicating a weak repulsive interaction. 
{If the data for large $k$ can be taken seriously,  there seems to be a flip of sign around scattering momentum $ak \approx 0.18$ and a rapid growth toward $pi/2$ beyond this point. Without further study, we cannot say whether this corresponds to a resonance or not.}

%then the sign flip in $\delta$ could correspond to a resonance around scattering momentum $ak \approx 0.18$. 

Finally, in Figure \ref{fig:phaseshiftminus} we show the results for the phase shift with $J^P= 0^-$.
In this channel there are no mixings, and one can expect a cleaner determination with respect to $J^P=0^+$.
Again, we neglect partial waves with $J>1$:
\begin{equation}
\cot \delta_{0^-} = \omega_{00}.
\end{equation}
The phase shift values are consistent with an attractive interaction.
$\delta$ increases with increasing $k$, which might indicate a resonance for { $ak>0.18$}.

\section{Summary and Outlook}

In this paper we have rederived the L\"uscher formalism for particles with general spin.
We find complete agreement with Ref.~\cite{Briceno:2014oea}.
We have explicitly formulated this approach for the case of two vector particles in the scalar channel.

The formalism is applied to scalar QED in the Higgs phase, where the gauge boson becomes massive.
For this model we derived the relevant operators to study scalar, vector and two vector particles with center of mass momenta up to $a\mathbf{p} = \frac{2\pi}{L}(1,1,1)$.
We have simulated scalar QED using Markov chains and {we have} estimated interacting and noninteracting energy levels for various total momenta in the scalar channel.
We have studied a set of bare parameter values and two volumes.
Even though the model is sufficiently simple to simulate, it is still a challenge to gather enough statistics to obtain significant results.
In general, the correlation functions measured by us, appear to be rather noisy.

In addition to the noise, it turns out that we are facing a dependence of the estimated single particle energy levels on the total momentum, which could be explained with lattice artifacts.
In the energy shift $\Delta E$, this dependence is much less pronounced.
However, statistical uncertainties are also {larger} for $\Delta E$.
Still, $\Delta E$ shows the expected dependence on $L$.
This makes us confident that our measurements are meaningful to a certain extent. Neglecting at this level any mixings and higher partial waves, we could extract the phase shift as a function of the scattering momentum.

A model-independent determination of all $S$-matrix parameters is computationally very expensive.
Namely, it would require many volumes and the use of the multichannel effective-range expansion (Eq.~(\ref{eq:EREMC})).
However, our results show that it is feasible to study the interaction of two vector particles.
Hence, in the future, we plan to apply these ideas to study the possibility of the Higgs boson to be a bound state of two $W$ bosons.

\begin{acknowledgments}

We would like to acknowledge the lattice group in Bonn and Roberto
Frezzotti for the interesting discussions and the support
provided. This work was supported in part by the DFG in the
Sino-German CRC110, by Volkswagenstiftung
under Contract No. 93562
and by Shota Rustaveli National Science Foundation (SRNSF), Grant No. DI-2016-26. Part of the computer time for this project was made available to us by Jureca in J\"ulich.
Finally, special thanks to BCGS for the continuous support.

\end{acknowledgments}

\begin{appendix}

\section{Convention for the Irreducible Representations \label{app:convention}}
\setcounter{table}{0}
\renewcommand{\thetable}{A\arabic{table}}
The ten irreducible representations (irreps) of the spatial symmetry groups of the lattice include 4 one-dimensional, 2 two-dimensional and 4 three-dimensional ones. They are: 
\begin{itemize}
 \item $A_1$ is the trivial representation, where all elements of $O_h$ are 1.
 \item $A_2$ is the trivial representation for $O$ times $-1$ when an inversion is present.
 \item $B_1$ assigns $R_i=-1$ to rotations in the conjugacy classes $6C_4$ and $6C_2'$ and $R=1$ otherwise.
 \item $B_2$ is the same as $B_1$ multiplying by $-1$ when an inversion is present.
 \item $E$ labels a two-dimensional representation. For the octahedral group, the superscript $E^\pm$ means whether an inversion multiplies the element by $\pm1$.%In Table \ref{tab:repE}, the elements of the two dimensional irreducible representation are shown for the difference spacial symmetry groups.
 \item $T_1^\pm$ is a three-dimensional representation which coincides with the Wigner matrices: $R_i = \exp (-i n^i J \omega_i)$, with $J$ the group generators and $n^i$ and $\omega_i$ as listed in Table \ref{tab:rotationO}. The superscript $\pm$ labels whether spatial inversion are assigned always $+1$ or $\pm1$.  
 \item $T_2^\pm$ is the same as $T_1$ with a change of sign in the conjugacy classes $6C_4$ and $6C_2'$.
\end{itemize}

In Appendix \ref{app:grouptables} the elements and characters of the different spatial symmetry groups are shown. They are taken to be in agreement with Ref. \cite{Gockeler:2012yj,Bernard:2008ax}.

\section{Group Tables \label{app:grouptables}}

\setcounter{table}{0}
\renewcommand{\thetable}{B\arabic{table}}

\begin{table}[H]
\centering
\begin{tabular}{|c|c|c|c|c|}
\hline
Class                    & $R_i$ & $\mathbf\ n$                        & $\omega$                     &   $R_i(E^\pm)$ \\ \hline
I                        & 1     & any                             & 0                            & $\mathbf{1}$ \\ \hline
\multirow{8}{*}{$8C_3$}  & 2     & $(1,1,1)$                       & $-2\pi/3$                    & $-\frac{1}{2} \mathbf{1} + \frac{i\sqrt{3}}{2} \sigma_2 $ \\ \cline{2-5} 
                         & 3     & $(1,1,1)$                       & $2\pi/3$                     & $-\frac{1}{2} \mathbf{1} - \frac{i\sqrt{3}}{2} \sigma_2 $\\ \cline{2-5} 
                         & 4     & $(-1,1,1)$                      & $-2\pi/3$                    & $-\frac{1}{2} \mathbf{1} - \frac{i\sqrt{3}}{2} \sigma_2 $\\ \cline{2-5}
                         & 5     & $(-1,1,1)$                      & $2\pi/3$                     & $-\frac{1}{2} \mathbf{1} + \frac{i\sqrt{3}}{2} \sigma_2 $ \\ \cline{2-5} 
                         & 6     & $(-1,-1,1)$                     & $-2\pi/3$                    & $-\frac{1}{2} \mathbf{1} + \frac{i\sqrt{3}}{2} \sigma_2 $ \\ \cline{2-5}
                         & 7     & $(-1,-1,1)$                     & $2\pi/3$                     & $-\frac{1}{2} \mathbf{1} - \frac{i\sqrt{3}}{2} \sigma_2 $\\ \cline{2-5}
                         & 8     & $(1,-1,1)$                      & $-2\pi/3$                    & $-\frac{1}{2} \mathbf{1} - \frac{i\sqrt{3}}{2} \sigma_2 $\\ \cline{2-5}
                         & 9     & $(1,-1,1)$                      & $2\pi/3$                     & $-\frac{1}{2} \mathbf{1} + \frac{i\sqrt{3}}{2} \sigma_2 $ \\ \cline{1-5}
\multirow{6}{*}{$6C_4$}  & 10    & $(1,0,0)$                       & $-\pi/2$                     & $-\frac{1}{2} \sigma_{3} - \frac{\sqrt{3}}{2} \sigma_1 $\\ \cline{2-5}
                         & 11    & $(1,0,0)$                       & $\pi/2$                      & $-\frac{1}{2} \sigma_{3} - \frac{\sqrt{3}}{2} \sigma_1 $\\ \cline{2-5}
                         & 12    & $(0,1,0)$                       & $-\pi/2$                     & $-\frac{1}{2} \sigma_{3} + \frac{\sqrt{3}}{2} \sigma_1 $\\ \cline{2-5}
                         & 13    & $(0,1,0)$                       & $\pi/2$                      & $-\frac{1}{2} \sigma_{3} + \frac{\sqrt{3}}{2} \sigma_1 $\\ \cline{2-5}
                         & 14    & $(0,0,1)$                       & $-\pi/2$                     & $\sigma_3$ \\ \cline{2-5} 
                         & 15    & $(0,0,1)$                       & $\pi/2$                      & $\sigma_3$ \\ \cline{1-5} 
\multirow{6}{*}{$6C'_2$} & 16    & $(0,1,1)$                       & $-\pi$                       & $-\frac{1}{2} \sigma_{3} - \frac{\sqrt{3}}{2} \sigma_1 $\\ \cline{2-5}
                         & 17    & $(0,-1,1)$                       & $-\pi$                       & $-\frac{1}{2} \sigma_{3} - \frac{\sqrt{3}}{2} \sigma_1 $\\ \cline{2-5}
                         & 18    & $(1,1,0)$                       & $-\pi$                       & $\sigma_3$ \\ \cline{2-5} 
                         & 19    & $(1,-1,0)$                      & $-\pi$                       & $\sigma_3$ \\ \cline{2-5} 
                         & 20    & $(1,0,1)$                       & $-\pi$                       & $-\frac{1}{2} \sigma_{3} + \frac{\sqrt{3}}{2} \sigma_1 $\\ \cline{2-5}
                         & 21    & $(-1,0,1)$                      & $-\pi$                       & $-\frac{1}{2} \sigma_{3} + \frac{\sqrt{3}}{2} \sigma_1 $\\ \cline{1-5}
\multirow{3}{*}{$3C_2$}  & 22    & $(1,0,0)$                       & $-\pi$                       & $\mathbf{1}$ \\ \cline{2-5} 
                         & 23    & $(0,1,0)$                       & $-\pi$                       & $\mathbf{1}$ \\ \cline{2-5} 
                         & 24    & $(0,0,1)$                       & $-\pi$                       & $\mathbf{1}$ \\ \hline
\end{tabular}
\caption{Rotations of the group $O$. Last column includes the element of the two dimensional irreducible representation for the cubic group. }
\label{tab:rotationO}
\end{table}

\begin{center}
\begin{table}[H]
\resizebox{1\textwidth}{!}{\begin{minipage}{\textwidth}
\centering
\begin{tabular}{| c | c | c | c |}
\hline
Group & boost   & name & Elements \\ \hline
\multirow{3}{*}{$O_h$} & (0,0,1) & $C_{4v}$ & $R_1,R_{14},R_{15},R_{24},IR_{18}, IR_{19}, IR_{22}, IR_{23}$\\ \cline{2-4}
                       & (1,1,0) & $C_{2v}$ & $R_1,R_{18},IR_{19}, IR_{24}$\\ \cline{2-4}
                       & (1,1,1) & $C_{3v}$ & $R_1,R_{2},R_{3}, IR_{17}, IR_{19}, IR_{21}$\\ \hline

\end{tabular}  
\caption{Elements of the Little Groups. \label{tab:littlegroups}}
\end{minipage} }
\end{table}
\end{center}

\begin{center}
\begin{table}[H]
\resizebox{1\textwidth}{!}{\begin{minipage}{\textwidth}
\centering
\begin{tabular}{| c | c | }
\hline
%LG  & Elements & Matrices \\ \hline
\multirow{2}{*}{$C_{4v}$} & ($R_1,R_{14},R_{15},IR_{18}, IR_{19}, IR_{22}, IR_{23},R_{24}$)\\ %\cline{2-2}
  &  ($\mathbf{1}$,$\frac{-i}{\sqrt{2}}(\sigma_1+\sigma_2)$,$\frac{i}{\sqrt{2}}(\sigma_1+\sigma_2)$,$\sigma_3$,$-\sigma_3$,$\frac{1}{\sqrt{2}}(\sigma_1-\sigma_2)$,$\frac{1}{\sqrt{2}}(-\sigma_1+\sigma_2)$,$-\mathbf{1}$)\\ \hline
\multirow{2}{*}{$C_{3v}$} & ($R_1,R_{2},R_{3}, IR_{17}, IR_{19}, IR_{21}$)\\ %\cline{2-2}
 & ($\mathbf{1}$,$-\frac{1}{2} \mathbf{1} +i\frac{\sqrt{3}}{2}\sigma_2$,$-\frac{1}{2} \mathbf{1} -i\frac{\sqrt{3}}{2}\sigma_2$,$\frac{1}{2} \sigma_3 +\frac{\sqrt{3}}{2}\sigma_1$,$- \sigma_3$,$\frac{1}{2} \sigma_3 -\frac{\sqrt{3}}{2}\sigma_1$)\\ \hline
\end{tabular}  
\caption{Elements of the two dimensional irrep in the Little Groups. \label{tab:littlegroupsE}}
\end{minipage} }
\end{table}
\end{center}

\begin{center}
\begin{table}[H]
\centering
\begin{tabular}{|c|c|c|c|c|c|c|c|c|c|c|}
\hline
      & $\mathbf{1}$ & $8C_3$ & $6C_4$ & $6C_2'$ & $3C_2$ & $I$  & $8IC_3$ & $6IC_4$ & $6IC_2'$ & $3IC_2$ \\ \hline
$A_1$ & 1            & 1      & 1      & 1       & 1      & 1  & 1       & 1       & 1        & 1       \\ \hline
$A_2$ & 1            & 1      & 1      & 1       & 1      & -1 & -1      & -1      & -1       & -1      \\ \hline
$B_1$ & 1            & 1      & -1     & -1      & 1      & 1  & 1       & -1      & -1       & 1       \\ \hline
$B_2$ & 1            & 1     & -1     & -1      & 1      & -1 & -1      & 1       & 1        & -1      \\ \hline
\end{tabular}%
\caption{Table of elements in the different one dimensional representations. Valid for all symmetry groups.
\label{tab:1Delements}}
\end{table}
\end{center}

\section{Basis Vectors \label{app:BV}}
\setcounter{table}{0}
\renewcommand{\thetable}{C\arabic{table}}

\begin{center}
%\begin{figure}[H]
\begin{minipage}[b]{0.4\linewidth}
\begin{table}[H]
\centering
\begin{tabular}{| c | c | c |}
\hline
$\Gamma$ &$J$& Basis vectors  \\ \hline
$A_2$               & 0 & $\ket{0,0}$ \\ \hline\hline
\multirow{3}{*}{$T^+_1$}& 1 & $-\frac{1}{\sqrt{2}}(\ket{1,1}-\ket{1,-1}) $  \\ \cline{2-3}
                        & 1 & $\ket{1,0} $  \\ \cline{2-3}
                        & 1 & $\frac{i}{\sqrt{2}}(\ket{1,1}+\ket{1,-1}) $  \\ \hline\hline
\multirow{3}{*}{$T^-_2$}& 2 & $\frac{1}{\sqrt{2}}(\ket{2,1}+\ket{2,-1}) $  \normalsize \\ \cline{2-3}
                        & 2 & $\frac{i}{\sqrt{2}}(\ket{2,1}-\ket{2,-1}) $   \\ \cline{2-3}
                        & 2 & $\frac{1}{\sqrt{2}}( \ket{2,-2}-\ket{2,2})$ \normalsize \\ \hline
\multirow{2}{*}{$E^-$}& 2 & $\frac{1}{\sqrt{2}}( \ket{2,-2}+\ket{2,2})$ \normalsize \\ \cline{2-3}
                    & 2 & $\ket{2,0} $  \normalsize \\ \hline
\end{tabular}  
\caption{Basis vectors for $O_h$ with $(-1)^J\neq(-1)^l$. \label{tab:BV1d}}
\end{table}
\end{minipage}
\hspace{0.2\linewidth}
\begin{minipage}[b]{0.4\linewidth}
\begin{table}[H]
\centering
\begin{tabular}{| c | c | c |}
\hline
$\Gamma$ &$J$& Basis vectors  \\ \hline
$A_1$               & 0 & $\ket{0,0}$ \\ \hline\hline
\multirow{3}{*}{$T^-_1$}& 1 & $-\frac{1}{\sqrt{2}}(\ket{1,1}-\ket{1,-1}) $  \\ \cline{2-3}
                        & 1 & $\ket{1,0} $ \\ \cline{2-3}
                        & 1 & $\frac{i}{\sqrt{2}}(\ket{1,1}+\ket{1,-1}) $   \\ \hline\hline
\multirow{3}{*}{$T^+_2$}& 2 & $\frac{1}{\sqrt{2}}(\ket{2,1}+\ket{2,-1}) $   \normalsize \\ \cline{2-3}
                        & 2 & $\frac{i}{\sqrt{2}}(\ket{2,1}-\ket{2,-1}) $   \\ \cline{2-3}
                        & 2 & $\frac{1}{\sqrt{2}}( \ket{2,-2}-\ket{2,2})$ \normalsize \\ \hline
\multirow{2}{*}{$E^+$}& 2 & $\frac{1}{\sqrt{2}}( \ket{2,-2}+\ket{2,2})$ \normalsize \\ \cline{2-3}
                    & 2 & $\ket{2,0} $  \normalsize \\ \hline
\end{tabular}  
\caption{Basis vectors for $O_h$ with $(-1)^J = (-1)^l$. \label{tab:BV2d}}
\end{table}
\end{minipage}
%\end{figure}
\end{center}

\begin{center}
\begin{figure}[H]
\begin{minipage}[b]{0.40\linewidth}
\begin{table}[H]
\centering
\begin{tabular}{| c | c | c | c |}
\hline
$\Gamma$ &$J$&$\alpha$ & Basis vectors  \\ \hline
$A_2$               & 0 &   & $\ket{0,0}$ \\ \hline\hline
$A_2$               & 1 &   & $\ket{1,0}$ \\ \hline
\multirow{2}{*}{$E$}& 1 & 1 & $\frac{1}{2}(1+i)\ket{1,-1} -\frac{1}{2}(1-i)\ket{1,1} $  \\ \cline{2-4}
                    & 1 & 2 & $\frac{1}{\sqrt{2}}\ket{1,-1} -\frac{1}{\sqrt{2}}i\ket{1,1} $  \\ \hline\hline
$A_2$               & 2 &   & $\ket{2,0}$ \\ \hline
$B_1$               & 2 &   & $\frac{1}{\sqrt{2}}( \ket{2,-2}-\ket{2,2})$  \\ \hline
$B_2$               & 2 &   & $\frac{1}{\sqrt{2}}( \ket{2,-2}+\ket{2,2})$  \\ \hline
\multirow{2}{*}{$E$}& 2 & 1 & $\frac{1}{2}(1-i)\ket{2,-1} -\frac{1}{2}(1+i)\ket{2,1} $  \\ \cline{2-4}
                    & 2 & 2 & $\frac{1}{\sqrt{2}}\ket{2,-1} +\frac{i}{\sqrt{2}}\ket{2,1} $  \\ \hline
\end{tabular}  
\caption{Basis vectors for $C_{4v}$ with $(-1)^J\neq(-1)^l$. \label{tab:BV1a}}
\end{table}
\end{minipage}
\hspace{0.2\linewidth}
\begin{minipage}[b]{0.40\linewidth}
\begin{table}[H]
\centering
\begin{tabular}{| c | c | c | c |}
\hline
$\Gamma$ &$J$&$\alpha$ & Basis vectors  \\ \hline
$A_1$               & 0 &   & $\ket{0,0}$ \\ \hline\hline
$A_1$               & 1 &   & $\ket{1,0}$ \\ \hline
\multirow{2}{*}{$E$}& 1 & 1 &$\frac{1}{2}(1+i)\ket{1,-1} +\frac{1}{2}(1-i)\ket{1,1} $  \\ \cline{2-4}
                    & 1 & 2 &$\frac{1}{\sqrt{2}}\ket{1,-1} +\frac{1}{\sqrt{2}}i\ket{1,1} $  \\ \hline\hline
$A_1$               & 2 &   & $\ket{2,0}$ \\ \hline
$B_1$               & 2 &   & $\frac{1}{\sqrt{2}}( \ket{2,-2}+\ket{2,2})$  \\ \hline
$B_2$               & 2 &   & $\frac{1}{\sqrt{2}}( \ket{2,-2}-\ket{2,2})$  \\ \hline
\multirow{2}{*}{$E$}& 2 & 1 & $\frac{1}{2}(1-i)\ket{2,-1} +\frac{1}{2}(1+i)\ket{2,1} $  \\ \cline{2-4}
                    & 2 & 2 & $\frac{1}{\sqrt{2}}\ket{2,-1} -\frac{i}{\sqrt{2}}\ket{2,1} $  \\ \hline
\end{tabular}  
\caption{Basis vectors for $C_{4v}$ with $(-1)^J = (-1)^l$. \label{tab:BV2a}}
\end{table}
\end{minipage}
\end{figure}
\end{center}

\begin{center}
\begin{figure}[H]
\begin{minipage}[b]{0.4\linewidth}
\begin{table}[H]
\centering
\begin{tabular}{| c | c | c | c |}
\hline
$\Gamma$ &$J$&$n_\Gamma$ & Basis vectors  \\ \hline
$A_2$               & 0 &   & $\ket{0,0}$ \\ \hline\hline
$A_2$               & 1 &   & $\frac{1}{\sqrt{2}}\ket{1,-1}+ \frac{i}{\sqrt{2}}\ket{1,1}$  \\ \hline
$B_2$               & 1 &   & $\frac{1}{\sqrt{2}}\ket{1,-1}- \frac{i}{\sqrt{2}}\ket{1,1}$ \normalsize \\ \hline
$B_1$               & 1 &   & $\ket{1,0} $\normalsize \\ \hline\hline
$A_2$               & 2 & 1 & $\ket{2,0}$ \\ \hline
$A_2$               & 2 & 2 & $\frac{1}{\sqrt{2}}( \ket{2,-2}-\ket{2,2})$ \normalsize \\ \hline
$A_1$               & 2 &   & $\frac{1}{\sqrt{2}}( \ket{2,-1}-i\ket{2,1})$ \normalsize \\ \hline
$B_2$               & 2 &   & $\frac{1}{\sqrt{2}}( \ket{2,-2}+\ket{2,2})$\normalsize \\ \hline
$B_1$               & 2 &   & $\frac{1}{\sqrt{2}}( \ket{2,-1}+i\ket{2,1})$ \normalsize \\ \hline
\end{tabular}  
\caption{Basis vectors for $C_{2v}$ with $(-1)^J\neq(-1)^l$. \label{tab:BV1b}}
\end{table}
\end{minipage}
\hspace{0.2\linewidth}
\begin{minipage}[b]{0.4\linewidth}
\begin{table}[H]
\centering
\begin{tabular}{| c | c | c | c |}
\hline
$\Gamma$ &$J$&$n_\Gamma$ & Basis vectors  \\ \hline
$A_1$               & 0 &   & $\ket{0,0}$ \\ \hline\hline
$A_1$               & 1 &   & $\frac{1}{\sqrt{2}}\ket{1,-1}+ \frac{i}{\sqrt{2}}\ket{1,1}$  \\ \hline
$B_1$               & 1 &   & $\frac{1}{\sqrt{2}}\ket{1,-1}- \frac{i}{\sqrt{2}}\ket{1,1}$ \normalsize \\ \hline
$B_2$               & 1 &   & $\ket{1,0} $\normalsize \\ \hline\hline
$A_1$               & 2 & 1 & $\ket{2,0}$ \\ \hline
$A_1$               & 2 & 2 & $\frac{1}{\sqrt{2}}( \ket{2,-2}-\ket{2,2})$ \normalsize \\ \hline
$A_2$               & 2 &   & $\frac{1}{\sqrt{2}}( \ket{2,-1}-i\ket{2,1})$ \normalsize \\ \hline
$B_1$               & 2 &   & $\frac{1}{\sqrt{2}}( \ket{2,-2}+\ket{2,2})$\normalsize \\ \hline
$B_2$               & 2 &   & $\frac{1}{\sqrt{2}}( \ket{2,-1}+i\ket{2,1})$ \normalsize \\ \hline
\end{tabular}  
\caption{Basis vectors for $C_{2v}$ with $(-1)^J = (-1)^l$.\label{tab:BV2b}}
\end{table}
\end{minipage}
\end{figure}
\end{center}

\begin{center}
\begin{figure}[H]
\begin{minipage}[b]{0.51\linewidth}
\begin{table}[H]
\centering
\begin{tabular}{| c | c | c |}
\hline
$\Gamma$ &$J$& Basis vectors  \\ \hline
$A_2$               & 0 & $\ket{0,0}$ \\ \hline\hline
$A_2$               & 1 & $\frac{1}{\sqrt{3}}\ket{1,-1}+\frac{1-i}{\sqrt{6}}\ket{1,0}+\frac{i}{\sqrt{3}}\ket{1,1}$ \\ \hline
\multirow{2}{*}{$E$}& 1 & $\frac{-i}{\sqrt{6}}\ket{1,-1}+\frac{1+i}{\sqrt{3}}\ket{1,0}+\frac{1}{\sqrt{6}}\ket{1,1}$ \normalsize \\ \cline{2-3}
                    & 1 & $\frac{1}{\sqrt{2}}\ket{1,-1} -\frac{i}{\sqrt{2}}\ket{1,1} $ \normalsize \\ \hline\hline
$A_2$               & 2 &  $\frac{1}{\sqrt{6}}\ket{2,-2}+\frac{1-i}{\sqrt{6}}\ket{2,-1}+\frac{1+i}{\sqrt{6}}\ket{2,1}-\frac{1}{\sqrt{6}}\ket{2,2}$ \\ \hline
\multirow{2}{*}{$E$}& 2 & $\ket{2,0} $ \normalsize \\ \cline{2-3}
                    & 2 & $\frac{1}{\sqrt{2}}( \ket{2,-2}+\ket{2,2})$ \normalsize \\ \hline
\multirow{2}{*}{$E$}& 2 & $\frac{1-i}{\sqrt{6}}\ket{2,-2}+\frac{i}{\sqrt{6}}\ket{2,-1}-\frac{1}{\sqrt{6}}\ket{2,1}+\frac{-1+i}{\sqrt{6}}\ket{2,2}$ \normalsize \\ \cline{2-3}
                    & 2 & $\frac{1}{\sqrt{2}}( \ket{2,-1}-i\ket{2,1})$  \normalsize \\ \hline
\end{tabular}  
\caption{Basis vectors for $C_{3v}$ with $(-1)^J\neq(-1)^l$. \label{tab:BV1c}}
\end{table}
\end{minipage}
\begin{minipage}[b]{0.51\linewidth}
\begin{table}[H]
\centering
\begin{tabular}{| c | c | c |}
\hline
$\Gamma$ &$J$& Basis vectors  \\ \hline
$A_1$               & 0 & $\ket{0,0}$ \\ \hline\hline
$A_1$               & 1 & ${\frac{1}{\sqrt{3}}\ket{1,-1}+\frac{1-i}{\sqrt{6}}\ket{1,0}
+\frac{i}{\sqrt{3}}\ket{1,1}}$ \\ \hline
\multirow{2}{*}{$E$}& 1 & $\frac{-i}{\sqrt{2}}\ket{1,-1} -\frac{1}{\sqrt{2}}\ket{1,1} $ \normalsize \\ \cline{2-3}
                    & 1 &  ${\frac{1}{\sqrt{6}}\ket{1,-1}
+\frac{-1+i}{\sqrt{3}}\ket{1,0}+\frac{i}{\sqrt{6}}\ket{1,1}}$ \normalsize \\ \hline\hline
$A_1$               & 2 &  $\frac{1}{\sqrt{6}}\ket{2,-2}+\frac{1-i}{\sqrt{6}}\ket{2,-1}+\frac{1+i}{\sqrt{6}}\ket{2,1}-\frac{1}{\sqrt{6}}\ket{2,2}$ \\ \hline
\multirow{2}{*}{$E$}& 2 & 
%$\frac{1}{2}( (-1-i)\ket{2,-1}+(-1+i)\ket{2,1})$ \normalsize \\ \cline{2-3}
${\frac{1}{\sqrt{2}}(\ket{2,-1}-i\ket{2,1})}$ \normalsize \\ \cline{2-3}
                    & 2 & 
%$\frac{1}{\sqrt{3}}(\ket{2,-2}+\frac{1-i}{2}\ket{2,-1}-\frac{1+i}{2}\ket{2,1}-\ket{2,2})$ \normalsize \\ \hline
${-\frac{1-i}{\sqrt{6}}\ket{2,-2}-\frac{i}{\sqrt{6}}\ket{2,-1}+\frac{1}{\sqrt{6}}\ket{2,1}+\frac{1-i}{\sqrt{6}}\ket{2,2}}$ \normalsize \\ \hline
\multirow{2}{*}{$E$}& 2 & $\frac{-1}{\sqrt{2}}( \ket{2,-2}+\ket{2,2})$ \normalsize \\ \cline{2-3}
                    & 2 & $\ket{2,0} $  \normalsize \\ \hline
\end{tabular}  
\caption{Basis vectors for $C_{3v}$ with $(-1)^J = (-1)^l$. \label{tab:BV2c}}
\end{table}
\end{minipage}
\end{figure}
\end{center}

\section{Tables for the Numerical Results\label{app:results}}
\setcounter{table}{0}
\renewcommand{\thetable}{D\arabic{table}}

\begin{center}
\begin{table}[H]
\centering
\begin{tabular}{| c | c | c |}
\hline
 Ensemble &  $am_V$ & $am_\phi$ \\ \hline
A12 & 0.1367(69) & 0.784(34) \\ \hline
B12 & 0.1346(43) & 0.617(28) \\ \hline
C12 & 0.1215(38) & 0.505(22) \\ \hline
D12 & 0.1166(44) & 0.468(19) \\ \hline
E12 & 0.0953(48) & 0.342(15) \\ \hline
A16 & 0.1466(16) & 0.728(13)  \\ \hline
\end{tabular}  
\caption{Mass of the scalar and vector particle for the different ensembles.}
\end{table}
\end{center}

\begin{center}
\begin{table}[H]
\centering
\begin{tabular}{| c | c | c |c | c |}
\hline
Ensemble &  $aE_{A_1}$ & $aE_{E}$ \\ \hline
A12 & {0.531(13)} & 0.5382(18)  \\ \hline
B12 & {0.535(15)} & 0.5357(26) \\ \hline
C12 & {0.528(16)} & 0.5225(26)  \\ \hline
D12 & {0.524(19)} & 0.5300(27)  \\ \hline
E12 & {0.528(17)} & 0.5210(68)  \\ \hline
\end{tabular}
\caption{Energy in the $A_1$/$E$ irreps of $d=(0,0,1)$ for the ensembles of Table \ref{tab:ensembles12}. \label{tab:massrep12}}
\end{table}
\end{center}

\begin{table}[H]
\centering
\begin{tabular}{| c | c | c | c | c | c |c |}
\hline
    &  $aE_{A_1}$ & $a\Delta E_{A_1}$&  $aE_{E^+}$ & $a\Delta E_{E^+}$&  $aE_{T^+_2}$ & $a\Delta E_{T^+_2}$ \\ \hline
\small A12 &\small 0.3046(53)&\small  0.031(14)&\small 0.3007(54)&\small 0.027(15)   &\small 0.3076(66)&\small 0.034(15) \\ \hline
\small B12 &\small 0.242(15) &\small -0.028(17)&\small 0.2679(48)&\small -0.0013(93) &\small 0.245(10) &\small -0.025(13) \\ \hline
\small C12 &\small 0.2136(71)&\small -0.024(10)&\small 0.2241(52)&\small -0.0187(98) &\small 0.2503(39)&\small 0.0002(77) \\ \hline
\small D12 &\small 0.1905(74)&\small -0.043(11)&\small 0.2096(58)&\small -0.0236(99) &\small 0.2231(41)&\small -0.0100(83) \\ \hline
\small E12 &\small 0.1641(87)&\small -0.019(10)&\small 0.174(11) &\small -0.026(12)  &\small 0.1457(86)&\small -0.0400(90) \\ \hline
\small A16 &\small 0.3010(17)&\small  0.0079(36)&\small 0.3031(20)&\small 0.0100(37)   &\small 0.3022(18)&\small 0.0091(36) \\ \hline
\end{tabular}  
\caption{Energy and energy shift for two particles.\label{tab:twoparticles12}}
\end{table}

\begin{center}
\begin{figure}[htp!]
\begin{minipage}[p]{0.98\linewidth}
\vfill
\begin{table}[H]
\centering
\begin{tabular}{| c | c | c | c | c | }
\hline
$\mathbf n$  & $\Gamma$ &  $aE$ &$am_V$  & $\sqrt{(am_V)^2 + (a\mathbf p)^2}$ \\ \hline
$(0,0,0)$ & $T^-_1$  &  \multicolumn{3}{c|}{0.1466(16)}    \\ \hline 
\multirow{2}{*}{$(0,0,1)$} & $A_1$  & { 0.4176(26)} & {0.1420(77)} &  \multirow{2}{*}{0.4192(6)}\\ \cline{2-4}
  & $E$ & 0.41797(80) & 0.1431(24) &\\ \hline
\multirow{3}{*}{$(1,1,0)$} & $A_1$  & {0.5724(27)} & {0.139(11)} & \multirow{3}{*}{0.5744(4)}\\ \cline{2-4}
  & $B_1$ &{0.5702(28)} & {0.129(12)} & \\ \cline{2-4}
  & $B_2$ & 0.5694(10) & 0.1257(45) & \\ \hline
\multirow{2}{*}{$(1,1,1)$} & $A_1$  &  {0.6925(27)}& {0.130(14)} &\multirow{2}{*}{0.6958(6)} \\ \cline{2-4}
  & $E$ & {0.6907(29)} & {0.120(17)} &\\ \hline
\multirow{2}{*}{$(0,0,2)$} & $A_1$  & {0.777(17)} & --- & \multirow{2}{*}{0.7990(3)} \\ \cline{2-4}
  & $E$ & 0.7832(13)& --- &\\ \hline
\end{tabular}  
\caption{One-particle mass in the multiple representations with L=16 using the continuum dispersion relation. The last column is the expected energy with the rest frame mass. Where the slot is empty, the determination has not been possible due to the precision.  \label{tab:rep1p}}
\end{table}
%\vfill
\end{minipage}
%\caption{One particle mass in the multiple representations with L=16. \label{tab:rep1p}}
\end{figure}
\end{center}

\begin{minipage}[b]{1\linewidth}
%\centering
\begin{table}[H]
\centering
\resizebox{0.95\textwidth}{!}{\begin{minipage}{\textwidth}
\centering
%\fontsize{10}{10}
\begin{tabular}{| c | c | c | c | c | c | c |}
\hline
%\fontsize{10}{10}
\small $\Gamma$ &  $aE$ & $a \Delta E$& $aE_{CM}$ & $\gamma$ &$ak$ & $aq$      \\   \hline
%\small$A_1$  & \small 0.5696(18) &\small 0.0038(28)&\small 0.4223(26)  &\small  1.3486(38) &\small 0.1526(23) &\small 0.3885(59) \\ \hline 
\small$A_1$  & \small 0.5696(18) &\small 0.0038(28)&\small 0.4126(25)  &\small  1.3805(40) &\small 0.1452(24) &\small 0.3698(61) \\ \hline 
%\small$A_1$  & \small 0.9127(19) &\small 0.0052(27)&\small 0.5469(35)  &\small 1.6689(71) &\small 0.2323(23) &\small 0.5914(59) \\ \hline %This is (0,0,0)+(0,0,2)
%\small$A_1$  & \small 0.8414(14) &\small 0.0136(18)&\small 0.4033(33)  &\small 2.086(13) &\small 0.1389(29) &\small 0.3538(74) \\ \hline %This is (0,0,1)+(0,0,1)
%\small$A_2$  & \small 0.5614(13) &\small 0.0009(25)&\small 0.4109(18)  &\small  1.3661(28) &\small 0.1445(20) &\small 0.3679(52)\\ \hline 
%\small$B_1$  & \small 0.5674(19) &\small 0.0069(28)&\small 0.4193(27)  &\small  1.3531(40) &\small 0.1504(24) &\small 0.3831(61) \\ \hline 
%\small$B_2$  & \small 0.5666(11) &\small 0.0061(23)&\small 0.4182(15)  &\small  1.3548(22) &\small 0.1497(18) &\small 0.3811(47) \\ \hline 
%\small$E  $  & \small 0.5772(56) &\small 0.0167(60)&\small 0.4328(76)  &\small  1.334(11)  &\small 0.1598(54) &\small 0.407(14) \\ \hline 
\small$A_2$  & \small 0.5614(13) &\small -0.0044(25)&\small 0.4012(18)  &\small  1.3993(31) &\small 0.1369(21) &\small 0.3487(55)\\ \hline 
\small$B_1$  & \small 0.5674(19) &\small 0.0016(28)&\small 0.4096(26)  &\small  1.3853(42) &\small 0.1430(25) &\small 0.3642(63) \\ \hline 
\small$B_2$  & \small 0.5666(11) &\small 0.0008(23)&\small 0.4084(16)  &\small  1.3874(23) &\small 0.1422(20) &\small 0.3620(51) \\ \hline 
\small$E  $  & \small {0.5835(14)} &\small {0.0177(26)}&\small {0.4316(18)}  &\small  {1.3523(24)}  &\small {0.1583(35)} &\small {0.4031(89)} \\ \hline 
%
% $T^+_1$  & 0.3931(25)  & 1.0011(64) & --- & $\delta_{122}=-0.4(2.4)$ & ---\\ \hline 
\end{tabular}  
%\caption{Phase shift for the irreps $A_1$ and  $T^+_1$. \label{tab:resdelta16}}
\caption{Energy for two particles in the moving frame with $d=(0,0,1)$. \label{tab:scatt001}}
\end{minipage} }
\end{table}
\end{minipage} 

\begin{minipage}[b]{1\linewidth}
%\centering
\begin{table}[H]
\centering
\resizebox{0.95\textwidth}{!}{\begin{minipage}{\textwidth}
\centering
%\fontsize{10}{10}
\begin{tabular}{| c | c | c | c | c |}
\hline
$L$ & Frame & $aE$ & $ak$ & $\delta_{0^+} \ (^o)$ \\ \hline 
\multirow{2}{*}{12}& $(0,0,0)+(0,0,0)$ &  0.3046(53)& 0.0582(82) & -2.09(52)  \\ \cline{2-5}
 & $(0,0,0)+(0,0,1)$ & 0.6840(40)& 0.1724(68)  & -0.9(1.2)\\ \hline
 \multirow{6}{*}{16}& $(0,0,0)+(0,0,0)$ & 0.3010(17)& 0.034(64) & -1.41(75) \\ \cline{2-5}
 & $(0,0,1)+(0,0,1)$ & 0.8414(14)& 0.036(11)&-0.80(72)\\ \cline{2-5}
 & $(0,0,0)+(0,0,1)$ & 0.5696(18) & 0.1452(24) & -1.13(76) \\ \cline{2-5}
 & $(0,0,0)+(1,1,0)$ & 0.7176(14) & 0.1736(20) & 1.60(83)\\ \cline{2-5}
 & $(0,0,0)+(0,0,2)$ & 0.9127(19) & 0.1804(27) &20.3(2.1) \\ \cline{2-5}
 & $(1,0,0)+(0,1,0)$ & 0.8361(22) & 0.2703(19) & 12.7(4.0)\\ \hline
% & $(0,0,0)+(1,1,1)$ & & &\\ \cline{2-5}
% & $(1,1,0)+(0,0,1)$ & & &\\ \cline{2-5}
\end{tabular}  
\caption{Energy spectrum in the $A_1$ representation. \label{tab:A1todo}}
\end{minipage} }
\end{table}
\end{minipage}

\begin{minipage}[b]{1\linewidth}
%\centering
\begin{table}[H]
\centering
\resizebox{0.95\textwidth}{!}{\begin{minipage}{\textwidth}
\centering
%\fontsize{10}{10}
\begin{tabular}{| c | c | c | c | c |}
\hline
$L$ & Frame & $aE$ & $ak$ & $\delta_{0^-} \ (^o)$ \\ \hline 
 \multirow{3}{*}{16}& $(0,0,0)+(0,0,1)$ & 0.5614(13)& 0.1369(21) & 1.2(7) \\ \cline{2-5}
 & $(0,0,0)+(1,1,0)$ & 0.7077(16)& 0.1631(23)&4.9(9)\\ \cline{2-5}
 & $(0,0,0)+(0,0,2)$ & 0.9101(18) & 0.1770(25) & 23(2)\\ \hline
% & $(0,0,0)+(1,1,1)$ & & &\\ \cline{2-5}
% & $(1,1,0)+(0,0,1)$ & & &\\ \cline{2-5}
\end{tabular}  
\caption{Energy spectrum in the $A_2$ representation. \label{tab:A2todo}}
\end{minipage} }
\end{table}
\end{minipage}

\begin{minipage}[b]{1\linewidth} 
\centering
\begin{table}[H]
\resizebox{\textwidth}{!}{\begin{minipage}{1\textwidth}
\centering
\begin{tabular}{| c | c | c | c |}
\hline
 $\Gamma$ & $ak$ & Formula &   $\delta \ (^o)$      \\   \hline
 $A_1$   &0.1452(24)& $\cot \delta_{1^-} = \omega_{00} - \omega_{20}$ & -0.14(10) \\ \hline 
 $A_2$   &0.1369(21)& $\cot \delta_{1^+} = \omega_{00} - \omega_{20}$ & 0.127(60)  \\ \hline 
\end{tabular}  
\caption{Obtained values for the phase shift $\delta_{1^\pm}$ in moving frame $(0,0,0)+(0,0,1)$ with the assumption of no mixings. \label{tab:phase001}}
\end{minipage} }
\end{table}
\end{minipage}

\section{Examples of $\mathcal{M}^\Gamma$ in the Rest Frame \label{app:M}}
\setcounter{table}{0}
\renewcommand{\thetable}{E\arabic{table}}

For the irreducible representations $E^+$ and $T_2^+$, with the notation:
\begin{equation}
 \mathcal{M}^{T_2^+/E^+}_{J'l',Jl} = 
 \begin{pmatrix}
(\mathcal{M}^{T_2^+/E^+}_{22,22})_{S=0} &  0 \\ 
0 &  (\mathcal{M}^{T_2^+/E^+}_{J'l',Jl})_{S=2} \\ 
 \end{pmatrix},
\end{equation}
with
\begin{equation}
 (\mathcal{M}^{T_2^+/E^+}_{J'l',Jl})_{S=2} = 
 \begin{pmatrix}
\mathcal{M}_{20,20} &  \mathcal{M}_{20,22} &  \mathcal{M}_{20,24} \\ 
\mathcal{M}_{22,20} &  \mathcal{M}_{22,22} &  \mathcal{M}_{22,24} \\ 
\mathcal{M}_{24,20} &  \mathcal{M}_{24,22} &  \mathcal{M}_{24,24} \\ 
 \end{pmatrix}.
\end{equation}
\begin{equation}
\mathcal{M}^{E^+} =
 \begin{pmatrix}
& \omega_{00}+\frac{18}{7} \omega_{40} & 0 & 0 & 0  \\
&0&\omega_{00}	&0&	3 \sqrt{\frac{2}{7}} \omega_{40} \\
&0&	0	&\omega_{00}+\frac{36}{49} \omega_{40}	&\frac{30\sqrt{5} }{49}  \omega_{40}\\
&0&	3 \sqrt{\frac{2}{7}} \omega_{40}&	\frac{30\sqrt{5}}{49}  \omega_{40}	&\omega_{00}+\frac{27}{49} \omega_{40}
 \end{pmatrix},
\end{equation}

\begin{equation}
\small
\mathcal{M}^{T_2^+} =
 \begin{pmatrix}
& \omega_{00}-\frac{12}{7} \omega_{40} & 0 & 0 & 0  \\
&0&\omega_{00}	&0&	-\frac{4}{\sqrt{14}}\omega_{40} \\
&0&	0	&\omega_{00}-\frac{24}{49} \omega_{40}	&-\frac{20\sqrt{5}}{49} \omega_{40}\\
&0&	-\frac{4}{\sqrt{14}}\omega_{40}&	-\frac{20\sqrt{5}}{49} \omega_{40}	&\omega_{00}-\frac{18}{49} \omega_{40} 
 \end{pmatrix}.
\end{equation}

For the representation $A_1$ with the notation
\begin{equation}
 \mathcal{M}^{A_1}_{J'l',Jl} = 
 \begin{pmatrix}
\mathcal{M}^{A_1}_{S=0} &  0 \\ 
0 &  \mathcal{M}^{A_1}_{S=2} \\ 
 \end{pmatrix},%= 
%\begin{pmatrix}
%\omega_{00} &  0 \\ 
%0 &  \omega_{00} \\ 
% \end{pmatrix}.
\end{equation}

\begin{equation}
 (\mathcal{M}^{A_1}_{J'l',Jl})_{S=0} = 
 \begin{pmatrix}
\mathcal{M}_{00,00} & \mathcal{M}_{00,44} \\ 
\mathcal{M}_{44,00} & \mathcal{M}_{44,44} \\ 
 \end{pmatrix},%= 
%\begin{pmatrix}
%\omega_{00} &  0 \\ 
%0 &  \omega_{00} \\ 
% \end{pmatrix}.
\end{equation}
\begin{equation}
 (\mathcal{M}^{A_1}_{J'l',Jl})_{S=2} = 
 \begin{pmatrix}
\mathcal{M}_{02,02} & \mathcal{M}_{02,42} & \mathcal{M}_{02,44} & \mathcal{M}_{02,46} \\ 
\mathcal{M}_{42,02} & \mathcal{M}_{42,42} & \mathcal{M}_{42,44} & \mathcal{M}_{42,46}\\ 
\mathcal{M}_{44,02} & \mathcal{M}_{44,42} & \mathcal{M}_{44,44} & \mathcal{M}_{44,46} \\ 
\mathcal{M}_{46,02} & \mathcal{M}_{46,42} & \mathcal{M}_{46,44} & \mathcal{M}_{46,46}\\ 
 \end{pmatrix},%= 
%\begin{pmatrix}
%\omega_{00} &  0 \\ 
%0 &  \omega_{00} \\ 
% \end{pmatrix}.
\end{equation}
\begin{equation}
 (\mathcal{M}^{A_1}_{J'l',Jl})_{S=0} = 
\begin{pmatrix}
\omega_{00} &  6 \sqrt{\frac{3}{7}} \omega_{40} \\ 
6 \sqrt{\frac{3}{7}} \omega_{40} &  \omega_{00}+\frac{4}{143} (81 \omega_{40}+260 \omega_{60}+140 \omega_{80}) \\ 
 \end{pmatrix}.
\end{equation}

\clearpage

\begin{sidewaysfigure*}
 
\begin{align}
 \begin{split}
(\mathcal{M}^{A_1}_{J'l',Jl})_{S=2} = 
\begin{pmatrix}
\omega_{00} &    \frac{6\sqrt{6}}{7} \omega_{40} & \frac{12}{7} \sqrt{\frac{15}{11}} \omega_{40} & 6 \sqrt{\frac{15}{77}} \omega_{40} \\ 
\frac{6\sqrt{6}}{7} \omega_{40} &  \omega_{00}+\frac{6}{7} \omega_{40} & \frac{12}{77} \sqrt{\frac{10}{11}} (9 \omega_{40}+35 \omega_{60}) & \frac{2}{143} \sqrt{\frac{70}{11}} (3 \omega_{40}+52 \omega_{60}+88 \omega_{80}) \\ 
\frac{12}{7} \sqrt{\frac{15}{11}} \omega_{40} & \frac{12}{77} \sqrt{\frac{10}{11}} (9 \omega_{40}+35 \omega_{60}) & \omega_{00}-\frac{2 (1863 \omega_{40}+20566 \omega_{60}-10976 \omega_{80})}{11011}& \frac{12 \sqrt{7} (45 \omega_{40}+208 \omega_{60}+72 \omega_{80})}{1573} \\
6 \sqrt{\frac{15}{77}} \omega_{40} & \frac{2}{143} \sqrt{\frac{70}{11}} (3 \omega_{40}+52 \omega_{60}+88 \omega_{80}) & \frac{12 \sqrt{7} (45 \omega_{40}+208 \omega_{60}+72 \omega_{80})}{1573} & \frac{1573 \omega_{00}+2352 \omega_{40}+4160 \omega_{60}+560 \omega_{80}}{1573} \\
 \end{pmatrix}.
 \end{split}
\end{align}
 
\end{sidewaysfigure*}

\clearpage

\section{Effective Mass \label{app:meff}}
\setcounter{table}{0}
\renewcommand{\thetable}{F\arabic{table}}

\begin{figure}[H]
   \centering
   \subfigure[Effective mass for the one-particle operator $\mathcal{O}^{T_1^-}$ in the rest frame as in Table \ref{tab:op1particle} \label{fig:meff16}]%
             {\includegraphics[width=0.475\textwidth,clip]{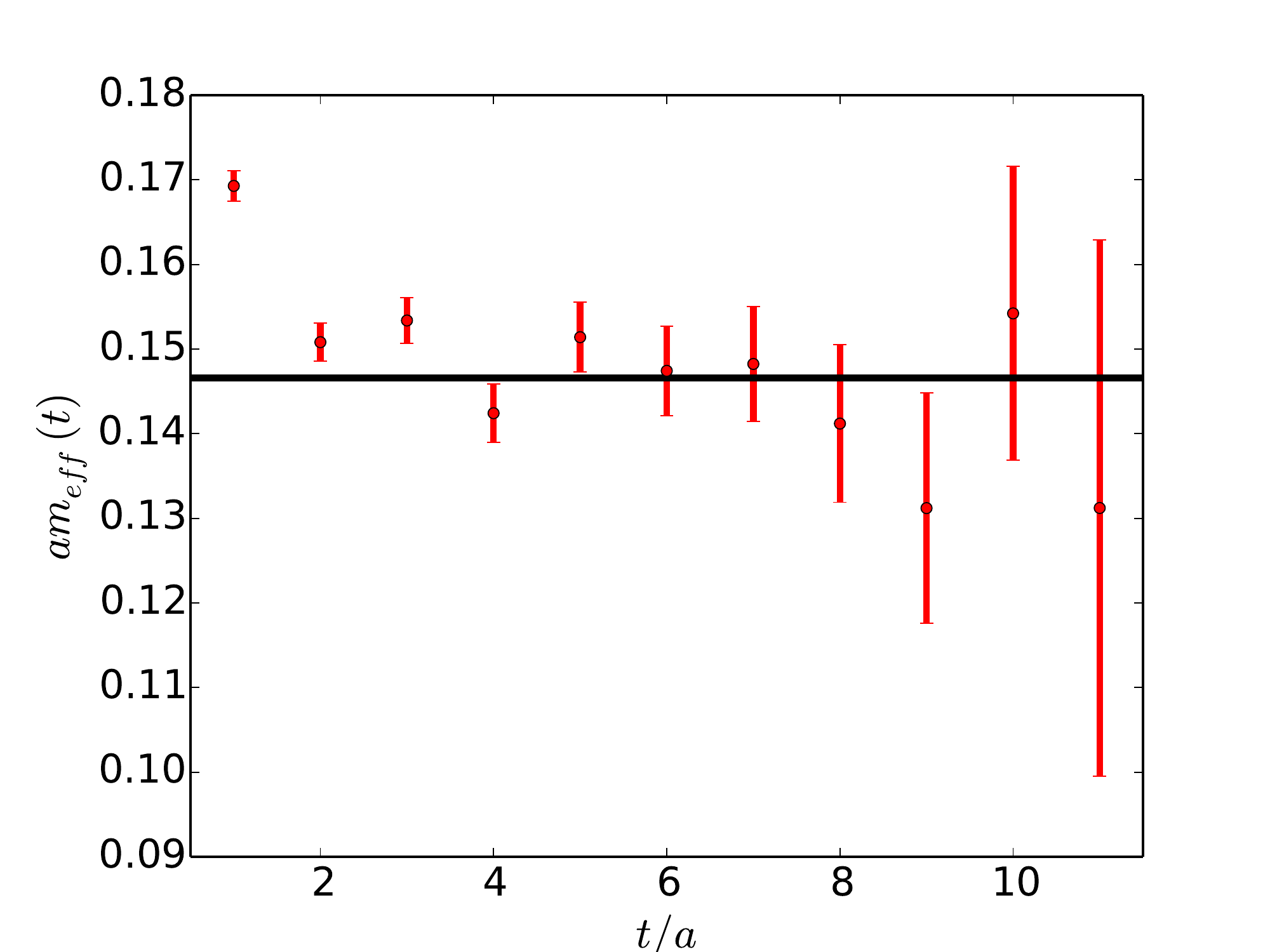}}\hfill   
   \subfigure[Effective mass for the two-particle operator $\mathcal{O}^{A_1}$ in the rest frame as in Table \ref{tab:op2particle} \label{fig:meffS0}]%
             {\includegraphics[width=0.475\textwidth,clip]{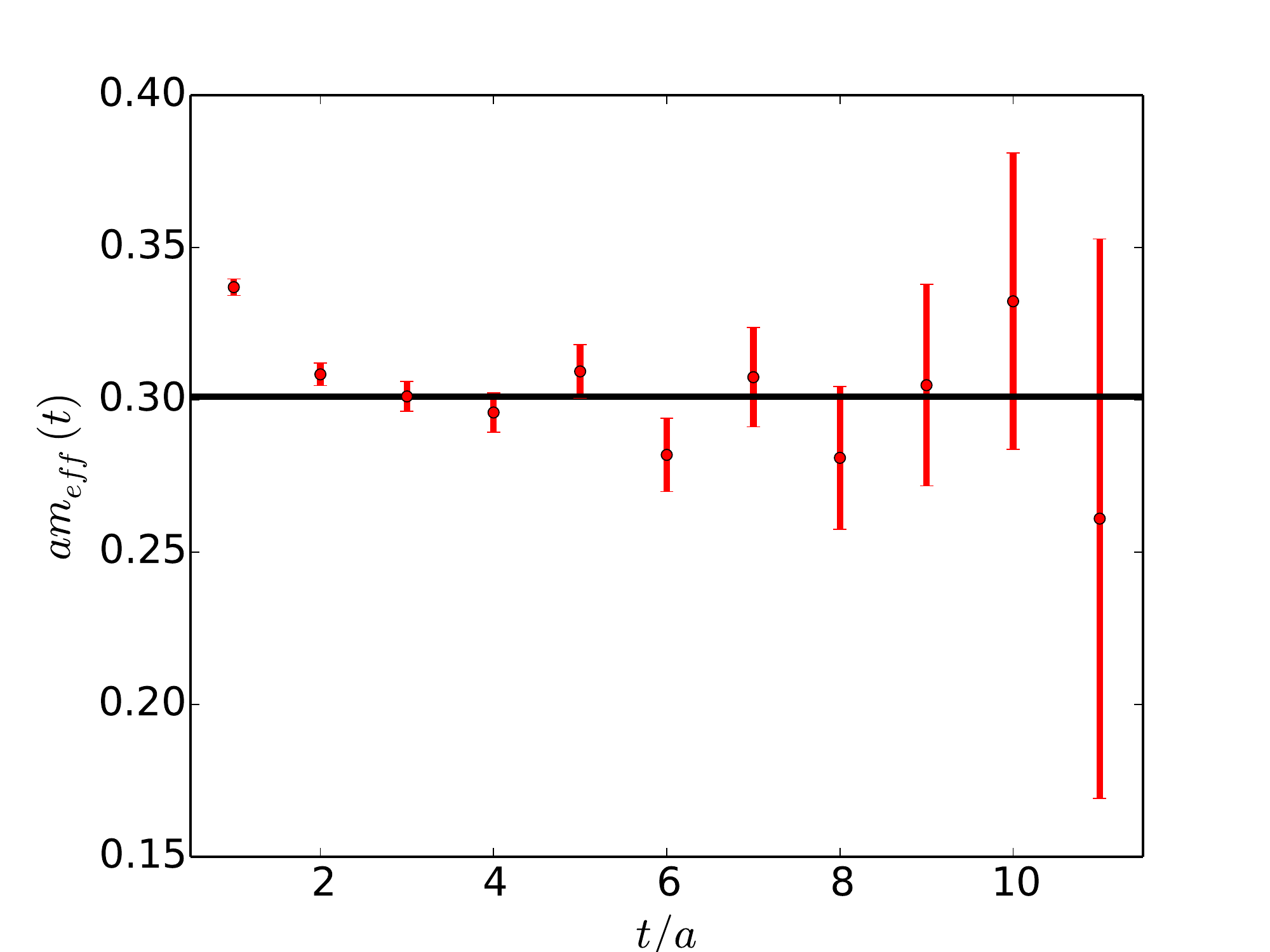}}
   \subfigure[{Effective mass for the one-particle operator $\mathcal{O}^{A_1}$ in the moving frame $ \mathbf d =(0,0,1)$ as in Table \ref{tab:op1particle} \label{fig:meffA1}}]%
             {\includegraphics[width=0.475\textwidth,clip]{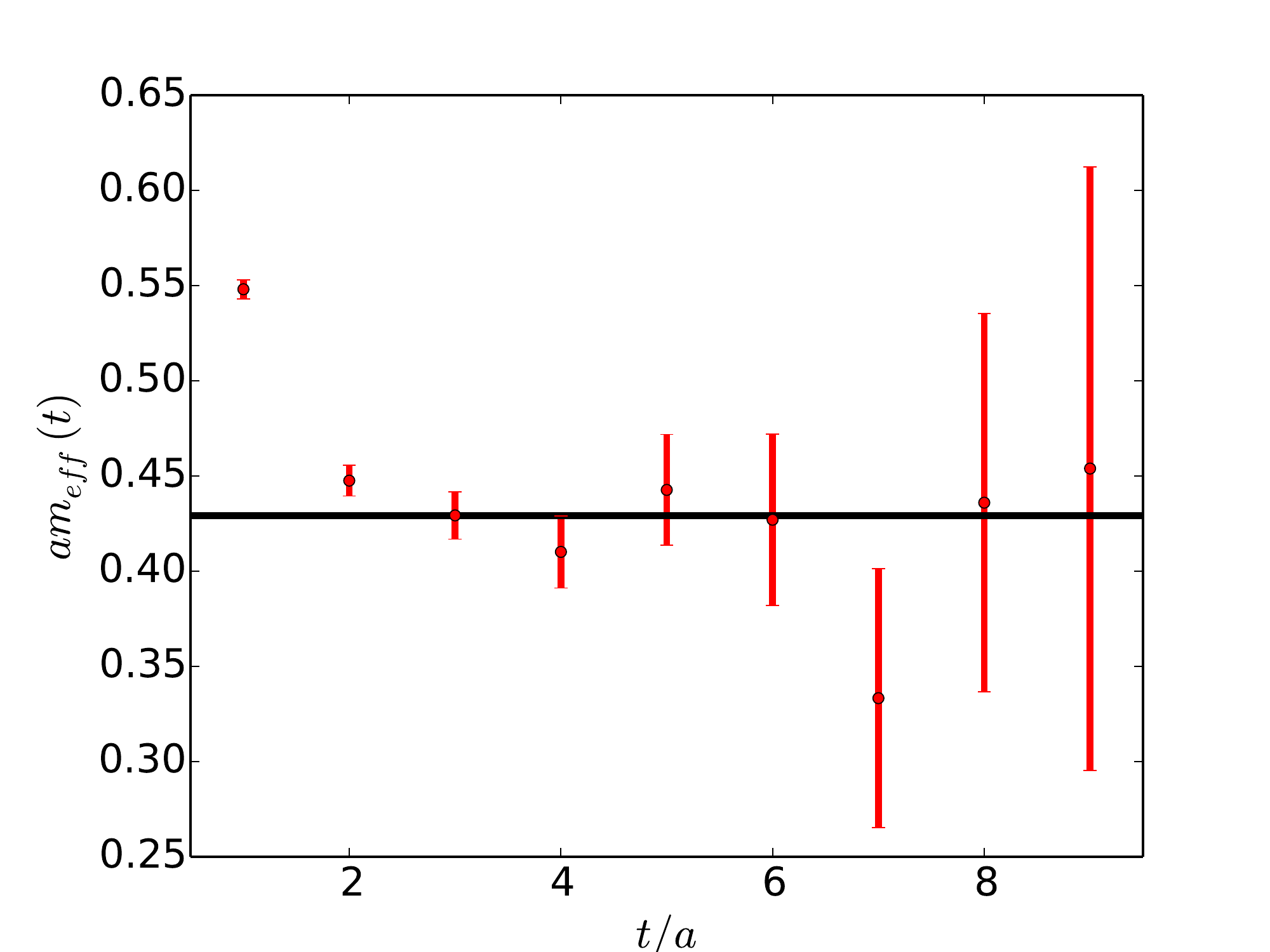}}\hfill   
   \subfigure[Effective mass for the two-particle operator $\mathcal{O}^{A_1}$ in the moving frame $ \mathbf d =(0,0,0)+(0,0,1)$ as in Table \ref{tab:op2particle} \label{fig:meffA12P}]%
             {\includegraphics[width=0.475\textwidth,clip]{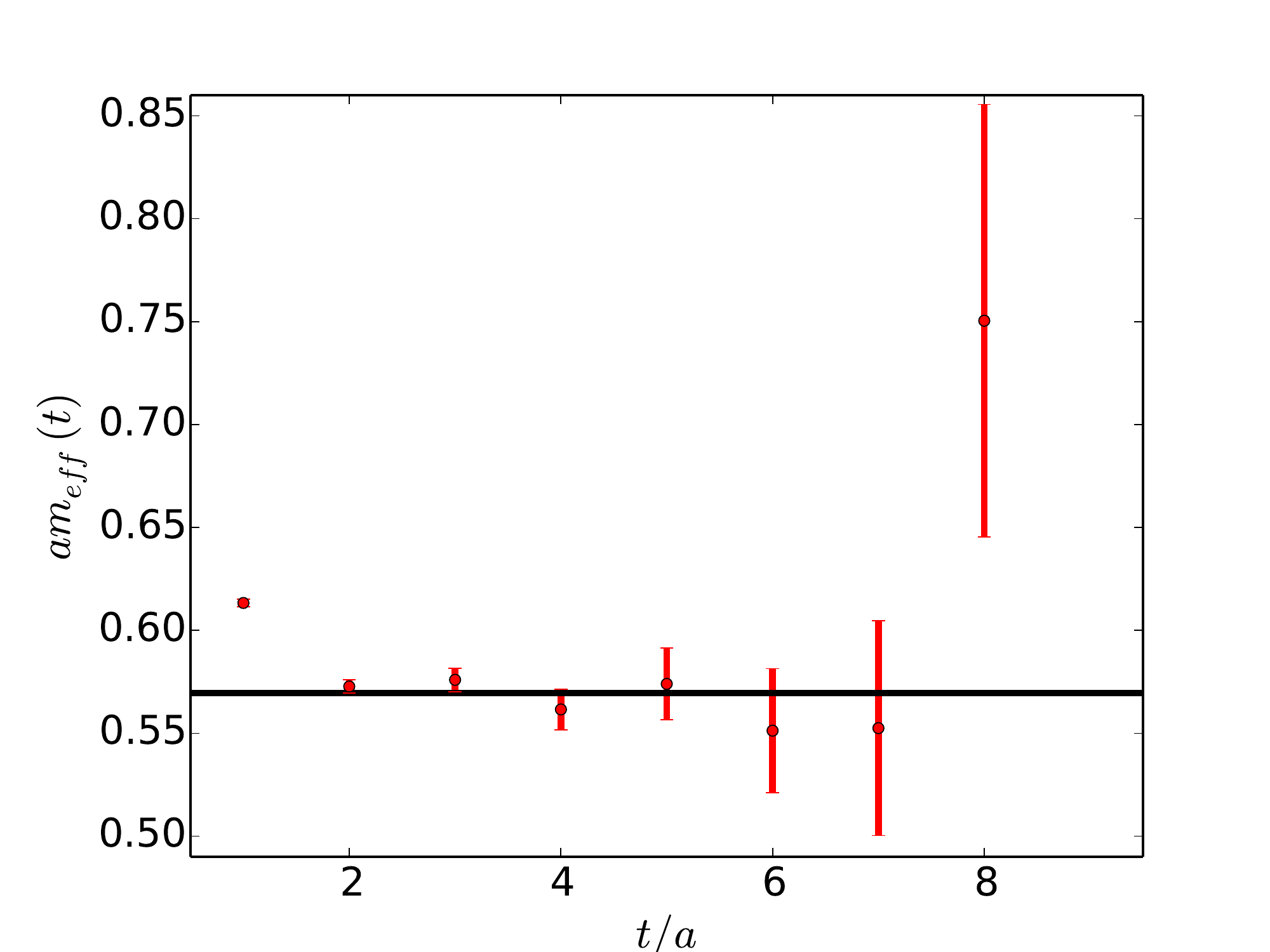}}
    \caption{Effective masses as a function of $t/a$ for ensemble A16}
   \label{fig:mefftot}
\end{figure}

\end{appendix}

\FloatBarrier
\bibliographystyle{h-physrev5}
\bibliography{lattice2017}

\end{document}